\documentclass[a4paper]{article}
\usepackage{amsfonts}
\usepackage{amsmath}
\usepackage{color}
\usepackage{float}
\usepackage{slashed}
\usepackage{subfigure}
\usepackage{pstricks}
\usepackage{graphicx}
\usepackage{epstopdf}
\usepackage{pstricks}
\newcommand{\be}{\begin{equation}}
\newcommand{\ee}{\end{equation}}
\newcommand{\ba}{\begin{eqnarray}}
\newcommand{\ea}{\end{eqnarray}}

\newcommand{\bea}{\begin{eqnarray}} \newcommand{\eea}{\end{eqnarray}}

\begin{document}
\title{Fermionic spectral functions in backreacting $p$-wave superconductors at finite temperature}
\author{
G.~L.~Giordano$^a$, N.~E.~Grandi$^a$, A.~R.~Lugo$^a$\\
{\normalsize $^a\!$ {\it Instituto de F\'i sica de La Plata - CONICET \& Departamento de F\' isica - UNLP,}}\\
{\normalsize {\it C.C. 67, 1900 La Plata, Argentina.}}\\
}
\date{\today}
\maketitle
\begin{abstract}
We investigate the spectral function of fermions in a $p$-wave superconducting state, at finite both temperature and gravitational coupling, using the $AdS/CFT$ correspondence and extending previous research. We found that, for any coupling  below a critical value, the system behaves as its zero temperature limit. By increasing the coupling, the ``peak-dip-hump" structure that characterizes the spectral function at fixed momenta disappears. In the region where the normal/superconductor phase transition is first order, the presence of a non-zero order parameter is reflected in the absence of rotational symmetry in the fermionic spectral function at the critical temperature.
\end{abstract}
\section{Introduction}
\label{sec:introduction}
One of the most relevant open problems in condensed matter theory is the understanding of high temperature superconductors. It is known that the standard BCS theory of low temperature superconductivity does not fit with the observed phenomenology, and that the correct description must take into account the high correlation of electrons in the fundamental state, that result in a strongly interacting Hamiltonian for the excitations. In such strongly interacting system, the usual perturbative treatment of quantum field theory does not apply, and the research has diversified into a bunch of alternative approaches \cite{Sachdev:2010un}.

In this context, an interesting issue is the description of the fermionic degrees of freedom, both in the normal and superconducting phases. These materials generically present a ``strange-metal'' phase which is not described by the standard Landau theory of the Fermi liquid (for example, the conductivity scales linearly with temperature, while Landau theory predicts a quadratic behavior).
Because of that, they are some times called ``non-Fermi liquids''.
The fermionic spectral function is studied by making use of ARPES spectroscopy \cite{Damascelli:2003uno}, uncovering a rich structure with
the Fermi surface evolving into Fermi arcs as temperature and/or doping are varied and a gap opens in some directions of momentum space.

Superconductors are classified according to the symmetry of their gap function \cite{Sigrist:1991uno}.
From this point of view, conventional BCS superconductors correspond to $s$-wave symmetry. Regarding unconventional high temperature superconductors, cuprates are believed to have $d$-wave symmetry, while some rutenates like Sr$_2$RuO$_4$ are supposed to have a $p$-wave gap.

~

Holographic superconductors \cite{HHH1}\cite{Hartnoll:2008kx}\cite{GubserP}\cite{Gubser:2010nc} are a class of theoretical models that share many defining features with high temperature superconductors. In particular, they are strongly coupled systems in which a superconducting phase arises at low temperatures. Moreover, the normal and superconducting states can propagate fermions that do not fit in the standard Landau description. The main difference with the laboratory systems is that, since holographic superconductors are described by means of the AdS/CFT correspondence, they require a large number of degrees of freedom at any spatial point, a condition referred as the ``large $N$ limit''.

The AdS/CFT correspondence \cite{Maldacena}\cite{Gubser:1998bc}\cite{Witten:1998qj} is a powerful tool to study strongly interacting systems. Roughly, it describes a $d$-dimensional quantum field theory in the strong coupling regime (or ``boundary theory'') in terms of a weakly coupled gravity theory in at least one dimension higher (or ``bulk theory''). The background of the bulk theory asymptotes Anti-de Sitter (AdS) space-time, while the boundary theory lives in the Minkowskian conformal boundary of AdS. Finite temperature states of the boundary theory are described by black hole backgrounds on the bulk side. On the other hand, finite chemical potential for a conserved particle number of the field theory is included by adding charge to the bulk solution.

In this setup, to model superconductors at finite temperature and chemical potential, a dynamical charged field describing the condensate needs to be added to the bulk theory. With this, we expect that for high temperatures the bulk background would be the AdS-Reissner-N\"ordstrom charged black hole dual to the normal phase, while for low temperatures the black hole would acquire a ``hair" of the condensate field, that extends non-trivially up to the boundary.  This is indeed what happens in the known holographic superconductor models.

For example, the $s$-wave holographic superconductor is modeled by adding a charged scalar field to the bulk, that is the dual to the scalar condensate.
Below a critical temperature a solution with non trivial scalar hair appears, which have less free energy that the AdS Reissner-N\"ordstrom  solution that describes the normal state.
This hairy black hole breaks the $U(1)$ gauge symmetry what, according to the AdS/CFT correspondence, is interpreted in the field theory as the spontaneous symmetry breaking of a $U(1)$ global symmetry due to the scalar condensate \cite{HHH1}\cite{Hartnoll:2008kx}\cite{Hartnoll:2009sz}\cite{Gubser1}\cite{Gubser:2008px}.

In this paper we are interested in $p$-wave holographic superconductors, where the condensate is some component of a vector field. In order for it to be charged, a non-Abelian gauge field is required.
In this system the normal phase is a solution with non trivial electric component in some direction of the gauge algebra,
representing the charge of the AdS Reissner-N\" ordstrom black hole.
We assume that this choice breaks the gauge group to a residual $U(1)$ gauge symmetry. Similarly to the $s$-wave case, the normal solution is energetically favored until a critical temperature is reached; below which a solution with a magnetic component in other gauge direction becomes non trivial, breaking spontaneously both the residual gauge $U(1)$ and the spatial rotational symmetries
\footnote{Higgs fields can be added but we will not do so here. Also isotropic solutions with  $\,A^1_{x^1}=A^2_{x^2}\neq 0\,$ that preserve the diagonal subgroup of $U(1)_{\it gauge}\times SO(2)_{\it rotation}$  can be obtained. However they always have free energy higher than the anisotropic solutions to be considered here, as showed explicitly in reference \cite{Giordano:2015vsa}.}. This is interpreted according to AdS/CFT as giving rise to a new phase in the boundary system, with a non trivial vector condensate that spontaneously breaks a global $U(1)$ symmetry \cite{GubserP}. The inclusion of the back-reaction of the gauge fields produces a change in the order of the phase transition at a critical gravity coupling, passing from second order to first order \cite{Ammon:2009xh}\cite{Gubser:2010dm}\cite{Nacho}\cite{HHV}.

Holographic fermions in the normal phase were studied in references \cite{Lee:2008xf}\cite{Liu:2009dm}\cite{Faulkner:2009wj}\cite{Cubrovic:2009ye}\cite{Faulkner:2011tm} among others, the conductivities for these systems were computed in \cite{Faulkner:2010da}\cite{Faulkner:2013bna}. The emerging picture is that of a Fermi liquid with a well defined Fermi surface, but without long-lived quasi-particles. Useful lectures on these subjects are \cite{Iqbal:2011ae}\cite{zaanenetal}. Regarding the superconducting phase, fermionic degrees of freedom on the $s$-wave superconductor were studied in \cite{Chen:2009pt}\cite{Basu:2010ak}. On the $p$-wave superconductor on the other hand, they were first  analyzed in \cite{Gubser:2010dm}, for the case of vanishing temperature at finite gravitational coupling. The case of finite temperature was studied in the probe limit in \cite{Ammon:2010pg}. It is the aim of this work to extend the computation to include both the effect of non-zero gravitational coupling and finite temperature \cite{Gubser:2010nc}.

~

The paper is divided as follows. In Section \ref{sec:background} we review the solutions that describe a $p$-wave superconductor, for vanishing and non-vanishing temperature. Section \ref{sec:adding-fermions} is devoted to the study of the equations of motion of fermions in the fundamental representation of $SU(2)$, the analysis of the boundary conditions, and the computation of the correlation functions following \cite{Iqbal:2009fd}\cite{Gubser:2010dm}. In Section \ref{sec:results} we present the numerical results, while that Section \ref{sec:discussion} is left to discussion and conclusions. Two appendices with some relevant results regarding the spectral function are added at the end.


\section{Bosonic sector: the holographic $p$-wave superconductor}
\label{sec:background}
In this section we establish our conventions and revisit the construction of solutions dual to holographic $p$-wave superconductors \cite{GubserP}, to be extensively used in rest of the paper. We first concentrate in the bulk solution, and then give the boundary interpretation.
\subsection{The bulk gravitational solution}
\label{sec:bulk}
Our starting point is a gravitational theory whose bosonic sector is a gravity-Yang-Mills system in a $3+1$ dimensional space-time, with Minkowskian metric $G_{MN}$ with $M,N=0,1,2,3$ and signature $(-+++)$, and gauge group $SU(2)$ with Hermitean generators $\tau^a$ with $a=0,1,2$. The action reads
\be
\label{eq:action-bosonic}
S^{(bos)}=
\int d^{4}x\,\sqrt{|G|}\; \left(\frac{1}{2\,\kappa^2}  \left(  R +\frac{6}{L^2} \right)
-\frac{1}{4\,e{}^2}\; F_{MN}^a F^{a\,MN}\right)\,.
\ee
Here $ \kappa^2\equiv 8\,\pi\, G_N$ with  $G_N$ the Newton constant, $L$ is the AdS scale related to the negative cosmological constant through $\Lambda\equiv -{3}/{L^2}$, $e$ is the gauge coupling and the gauge field strength is defined by $F^a_{MN} \equiv \partial_M A_N^a - \partial_N A_M^a + \epsilon_{abc}\;A_M^b\; A_N^c\,$. The equations of motion derived from this action read
\bea
\label{eq:eom-bosonic}
R_{MN}-\frac{R}{2}\; G_{MN}+\Lambda\, G_{MN}&=&
\frac{\kappa^2}{e^{2}}\;\left( F^{a}_{SM}F^{aS}_{N} - \frac{G_{MN}}{4}\; F^{a}_{PQ}\,F^{aPQ}\right)\,,
\cr
\nabla_{S}F^{a}_{MN}+\epsilon_{abc}\,A^{b}_{S}\,F^{c}_{MN}
&=&0\,,
\eea
where ``$\nabla$" stands for the usual geometric covariant derivative.

We are interested in solutions of the form \cite{GubserP}\cite{Ammon:2009xh}\cite{HHV}\cite{Giordano:2015vsa}
\bea
\label{eq:ansatz}
G &=& \frac{L^{2}}{z^2}\,\left(-f(z)\; s(z)^2\; dt^2 +\;\frac{ dz^2}{f(z)} + \frac{d x^{2}}{g(z)^2}
+ g(z)^{2}\,  d y^{2}\right)\,,\cr
{\bf A} &=& \phi(z)\;dt\;\tau_0 + W(z)\; dx\;\tau_1\,.
\eea
To understand this ansatz, we first observe that a non-vanishing $\phi(z)$ breaks the $SU(2)$ gauge symmetry down to $U(1)_{\tau_0}$, and selects a preferred time direction $t$. Then, a non-vanishing ``magnetic'' component $W(z)$ breaks the residual $U(1)_{\tau_0}$ completely, and establishes a spatial anisotropy between $x$ and $y$. Then if we take into account back-reaction effects the system cannot support a rotationally invariant metric, and that is why we need to include not only $f(z)$ and $s(z)$ but also an eventually non-trivial $g(z)$.
Replacing into the equations (\ref{eq:eom-bosonic}) we get
\bea
\label{eq:eom-anzatz}
s\,\left( \frac{\phi^{\prime}}{s}\right)^{\prime}&=&\frac{g^{2}\,W^2}{f}\;\phi\,,
\cr
\left(s\,f\,g^{2}\,W'\right)^{\prime}&=&-\frac{g^2\,\phi^{2}}{f\,s}\,W\,,
\cr
\frac{z^3}{s}\left(\frac{s\,f}{z^3}\right)^{\prime}&=&-\frac{3}{z}+\frac{\lambda^2}{2}\,\frac{z^3\,\phi'^2}{s^2}\,,
\cr
s^{\prime}&=&-z\,s\,\frac{g'^2}{g^2}-\frac{\lambda^{2}}{2}\,z^{3}\,g^{2}\,
\left(s\,W'^2 +\frac{\phi^{2}}{s\,f^{2}}\,W^2\right)\,,
\cr
g''-\frac{g'^2}{g}&=&
-\frac{g'}{z}\,\left(1-\frac{3}{f} +\frac{\lambda^2}{2}\,\frac{z^4\,\phi'^2}{f\,s^2}\right)
+\frac{\lambda^2}{2}\,z^2\,g^3\,\left( W'^2 -\frac{\phi^2\,W^2}{s^2\,f^2}\right)\,,
\eea
where we have defined the dimensionless coupling $\lambda \equiv \kappa/(L\,e)$
\footnote{The coupling constant $g_{YM}$ in \cite{Gubser:2010dm} and ours are related by, $\,\lambda=1/(\sqrt{2}\,g_{YM})$.}.

Notice that the ansatz \eqref{eq:ansatz} and equations of motion (\ref{eq:eom-anzatz}) are invariant under the scaling symmetries
\bea\label{eq:scale-symmetry}
(t\,\,;\,\, s, \phi) &\longrightarrow &\left(\frac{t}{\alpha}\,\,;\,\,\alpha s,\alpha \phi\right)\,,
\cr
(x, y\,\,;\,\,g,W) &\longrightarrow& \left(\frac{x}{\beta}, \beta y\,\,;\,\,\frac{g}{\beta}, \beta W\right)\;,
\cr
(x^\mu, z\,\,;\,\,\phi,W) &\longrightarrow & \left(\frac{1}{\gamma}x^\mu,\frac{1}{\gamma} z\,\,;\,\,\gamma \phi, \gamma W \right)\;.
\eea
These symmetries will be useful in what follows in order to fix the asymptotic values  of some of the fields.

The expressions \eqref{eq:eom-anzatz} represent a system of second order ordinary differential equations on the independent variable $z$, which moves in a certain interval $z\in [z_{UV},z_{IR}]$. It must be solved with boundary conditions at the frontiers $z_{UV}$ and $z_{IR}$ reflecting the desired physical situation. In what follows, we define $z_{UV}=0$ as the boundary of AdS. Close to $z_{UV}$, the above equations can be expanded in powers of $z$ and solved, to get the asymptotic behaviors
\bea\label{eq:boundary-UV}
\phi(z)&=& \mu +\rho\, z + \ldots\,,\cr
W(z)&=& W_{UV} + W'_{UV}\, z + \ldots\,,\cr
f(z)&=& 1 + f_{3}\, z^{3} + \ldots\,,\cr
s(z)&=& 1 - \frac{\lambda^{2}\,{W'}_{UV}^2}{8}\,z^{4}+ \ldots\,,\cr
g(z)&=& 1 + g_{3}\,z^{3} \ldots\,,
\eea
where we used the first two symmetries in (\ref{eq:scale-symmetry}) to fix the normalization conditions $s(0)= g(0)= 1$, in such a way to identify the $x^\mu$'s as Minkowskian coordinates at infinity. The third symmetry will be used differently according to whether we are studying the zero or finite Beckenstein-Hawking entropy cases in the forthcoming sections.

Now, by taking $\mu\neq0$ the boundary condition breaks explicitly the $SU(2)$ gauge symmetry to a $U(1)_{\tau_0}$ subgroup. Next, putting $W_{UV} =0$, implies that there is no explicit breaking of the residual $U(1)_{\tau_0}$ gauge symmetry. Then, a non-vanishing value of the quantity $W'_{UV}$ signals a spontaneous breaking of the remaining gauge $U(1)_{\tau_0}$ and rotational symmetries.

On the opposite extreme of the geometry,  the value of $z_{IR}$ and the details of the asymptotic expansions depend on whether we want
to study a solution with zero or finite entropy. In any case the Hawking temperature of the solution is determined by
\be\label{eq:T}
T = \frac{s(z_{IR})\,|f'(z_{IR})|}{4\,\pi} \equiv \frac{s_{IR}\,|f'_{IR}|}{4\,\pi}
\ee
\subsubsection{Vanishing Beckenstein-Hawking entropy}
\label{sec:bulk-zero-entropy}
In this case, we extend the geometry to the $z_{IR}\to\infty$ region to set our boundary conditions there. In such region, the equations of motion dictate the following behaviors
\bea
\label{eq:boundary-IR-zero-entropy}
\phi(z)&=&\phi_{IR}\, e^{-g_{IR}\,w_{IR}\,z}+\dots\,,\cr
W(z)&=& w_{IR}-\frac{\phi_{IR}^2}{4\,s^{2}_{0}\, g_{IR}^2\,w_{IR}}e^{-2\,g_{IR}\,w_{IR}\,z}+\dots\,,\cr
f(z)&=& 1-\frac{\lambda^{2}}{2}\,\frac{\phi_{IR}^{2}\,g_{IR}\,w_{IR}}{s_{IR}^2}\,z^{3}\,e^{-2\,g_{IR}\,w_{IR}\,z}+\dots\,,\cr
s(z)&=& s_{IR}+\frac{\lambda^{2}}{4}\,\frac{\phi_{IR}^2\,g_{IR}\,w_{IR}}{s_{IR}}\,z^{3}\,e^{-2\,g_{IR}\,w_{IR}\,z}+\dots\,,\cr
g(z)&=&g_{IR}-\frac{\lambda^{2}}{8}\,\frac{g_{IR}\,\phi_{IR}^2}{s_{IR}^2}\,z^2\,e^{-2\,g_{IR}\,w_{IR}\,z}+\dots\;.
\eea
In the above expressions we imposed smoothness conditions by putting to zero the coefficient of the divergent part of the solution to the linearised system. Generically, solutions with boundary conditions \eqref{eq:boundary-UV} and \eqref{eq:boundary-IR-zero-entropy} are domain walls that interpolate between two AdS spaces with equal scales and different light velocities, and describe the true vacuum of the theory. According to (\ref{eq:T}), other than zero entropy they also have zero temperature. Such solutions were first studied in \cite{Gubser:2010dm}\cite{Basu:2009vv} and revisited more recently in \cite{Giordano:2015vsa}. We fix $\mu=1$ by using the last scaling symmetry in (\ref{eq:scale-symmetry}).

\begin{figure}[H]
\begin{center}
\includegraphics[height=5cm,width=7cm]{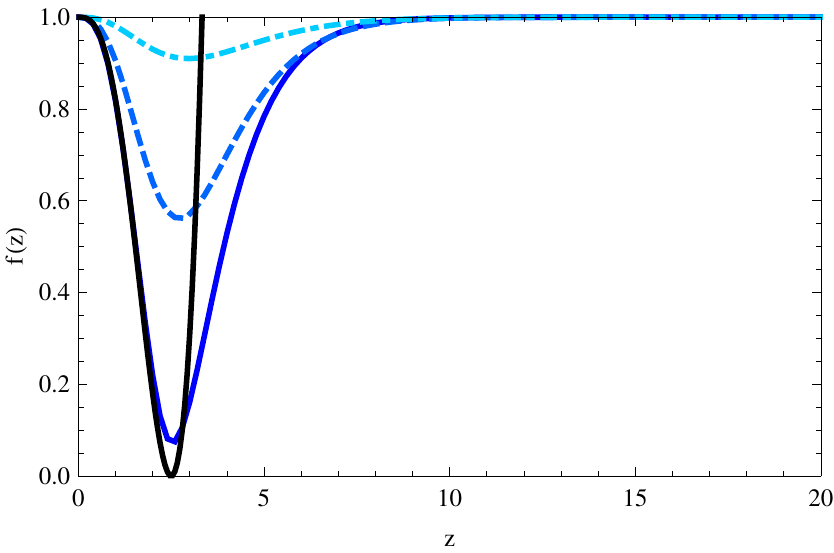}
\includegraphics[height=5cm,width=7cm]{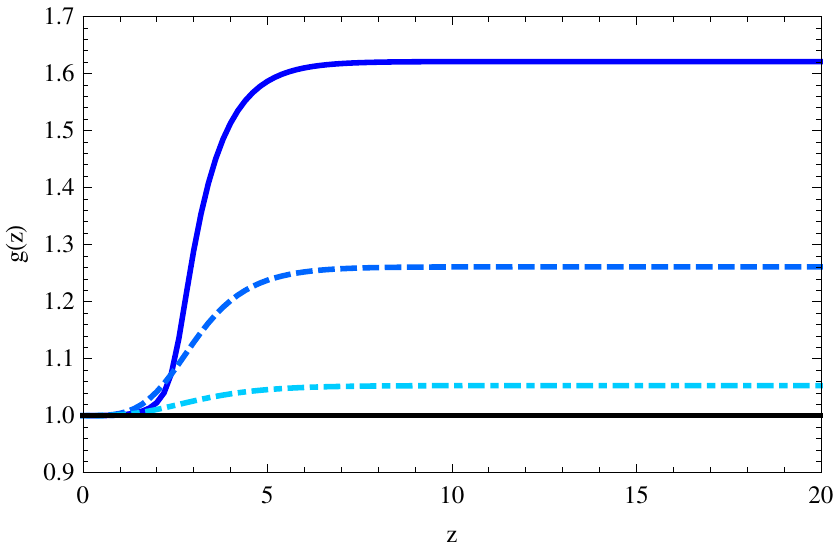}
\\
\includegraphics[height=5cm,width=7cm]{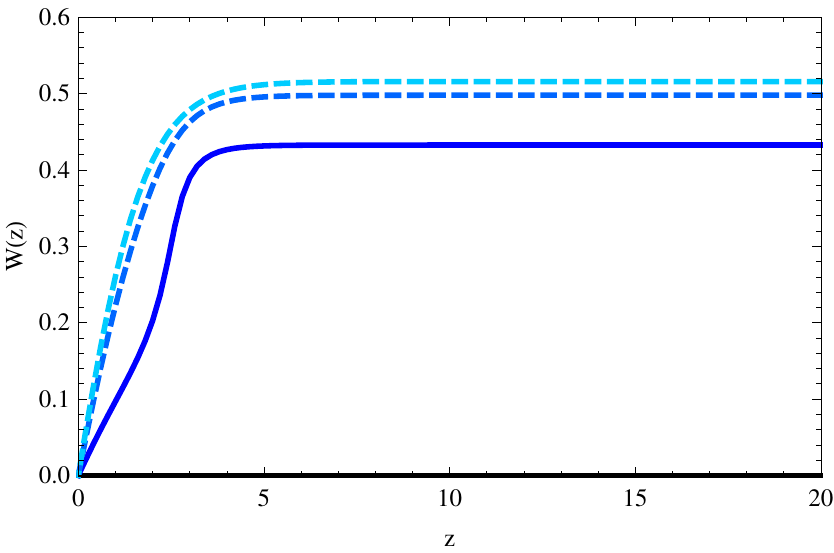}
\includegraphics[height=5cm,width=7cm]{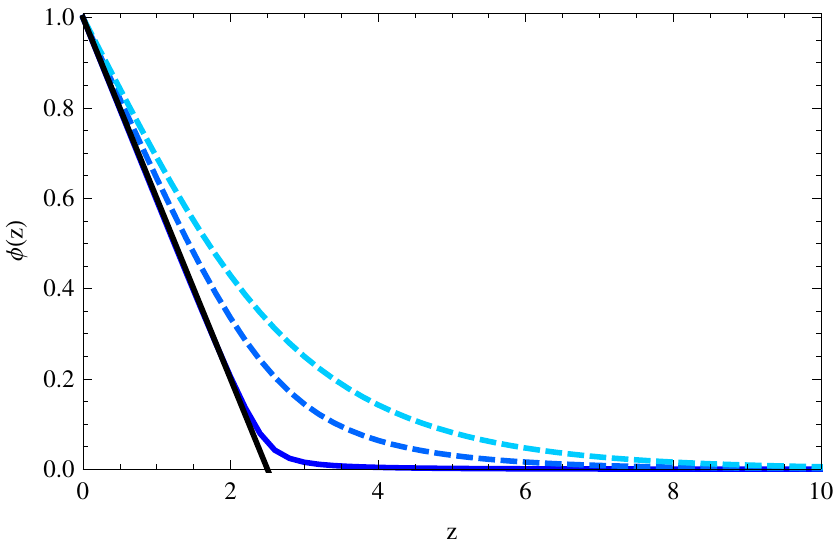}
\\
\includegraphics[height=5cm,width=7cm]{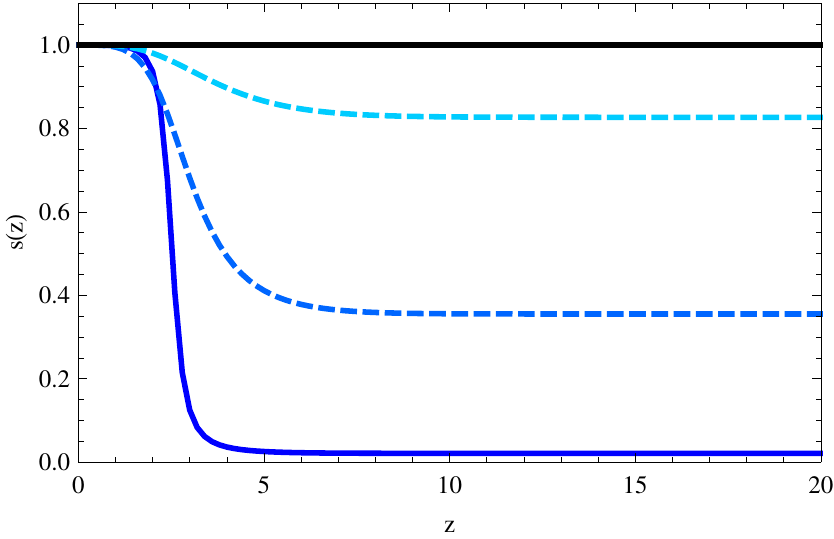}
\caption{
The metric and gauge field functions at null entropy, for three different values of $\lambda$: $\,0.40\,$ (cyan-blue), $0.8\,$ (light-blue), $0.995\,$ (blue).
The black curves correspond to the extremal Reissner-Nordstr\"om black hole solution.
\label{fig:functions-zero-entropy}
}
\end{center}
\end{figure}

In figure \ref{fig:functions-zero-entropy} the functions $f(z)$, $g(z)$, $W(z)$, $\phi(z)$ and $s(z)$ corresponding to different values of the coupling $\lambda$ are displayed.
We see that rotational and gauge symmetries are broken by the functions $g(z)$ and $W(z)$.
The linear behavior of $W(z)$ at the boundary signals the breaking of the $U(1)_{\tau_0}$ remaining gauge symmetry. When $\lambda\rightarrow\lambda_*\approx  0.995$ (corresponding to $g_{YM*}\approx 0.71$) a  horizon at $z_h\approx 2.462$ forms and the solution tends to the extremal AdS-Reissner-N\"ordstrom black hole
\footnote{
Strictly speaking, the AdS-Reissner-N\"ordstrom black hole is defined for $0\leq z\leq z_h$, see next subsection.
}
\bea
\label{eq:RN-extremal}
W(z)&=&0\,,
\cr
g(z)&=&s(z)\,\,=\,\,1\,,
\cr
\phi(z)&=&1 -\frac{z}{z_h}\,,
\cr
f(z) &=&  \left(1-\frac{z}{z_h}\right)^2\;
\left(1+2\,\frac{z}{z_h}+3\,\left(\frac{z}{z_h}\right)^2\right)\;.
\eea
For couplings above this quantum critical value $\lambda_*$ only the
AdS-Reissner-N\"ordstrom solution is present.
%

\subsubsection{Non-vanishing Beckenstein-Hawking entropy}
\label{sec:bulk-finite-entropy}
Now we consider solutions with a horizon at  $z_{IR}=z_h<\infty$ such that $f(z_h)\equiv0$.
The asymptotic expansion of the well-behaved solution near the horizon reads
\bea\label{eq:boundary-IR-finite-entropy}
\phi(z)&=&  \phi_1 (z - z_h) +\dots
\,,\cr
W(z)&=&  W_{IR} + W'_{IR}\, (z - z_h) +\dots
\,,\cr
f(z)&=& f'_{IR}\,(z-z_h) +\dots 
\,,\cr
s(z)&=& s_{IR} + s'_{IR}\,(z-z_h) + \dots 
\,,\cr
g(z)&=& g_{IR} + g'_{IR}\,(z-z_h) +\dots 
\,.
\eea
The boundary conditions \eqref{eq:boundary-UV} and \eqref{eq:boundary-IR-finite-entropy} are compatible with the AdS-Reissner-N\"ordstrom black hole solution
\bea
\label{eq:RN-nonextremal}
W(z)&=&0\,,
\cr
g(z)&=&s(z)\,\,=\,\,1\,,
\cr
\phi(z)&=&\mu\,\left(1 -\frac{z}{z_h}\right)\,,
\cr
f(z)&=&1-\left(1+Q^{2}\right)\,\left(\frac{z}{z_h}\right)^3+Q^{2}\,\left(\frac{z}{z_h}\right)^4\,,
\eea
where $Q^2\equiv {\lambda^2\,\mu^2\,z_h{}^2}/{2}$. The Hawking temperature of this black-hole solution is given by
\be
T = \frac{3-Q^{2}}{4\,\pi\, z_h}\,.
\label{eq:BHtemp}
\ee
The zero temperature limit corresponds to the extremal (finite entropy) AdS-Reissner-N\"ordstrom black hole with $Q^2=3$ considered above in equation (\ref{eq:RN-extremal}).
In fact, the extremal condition implies that the relation $\lambda_*\,z_h = \sqrt{6}$ should hold, consistent with the values numerically found, $\lambda_*\approx  0.995$ and $z_h\approx 2.462$.

Conditions \eqref{eq:boundary-UV} and \eqref{eq:boundary-IR-finite-entropy} are also compatible with a hairy ``magnetic'' $W(z)\neq0$ solution, that breaks the residual $U(1)_{\tau_0}$ gauge symmetry. To obtain it, we carry out numerical computations where, by using the last scaling symmetry in (\ref{eq:scale-symmetry}), we fix the position of the horizon at the value, $z_h=1$ \cite{GubserP}\cite{Ammon:2009xh}.

In terms of the scaling invariant temperature $\,T/\mu$, one gets that the AdS-Reissner-N\"ordstrom black hole solution is energetically favored for temperatures above a critical temperature $T_{c}$. For lower temperatures, the non symmetric hairy solution is selected.
As discussed for different models \cite{Gubser1}\cite{LMS2}\cite{LMS1} one can interpret this result by stating that a condensate is formed above a black hole horizon because of a balance of gravitational and electrostatic forces.
\subsection{The boundary QFT interpretation}
The $SU(2)$ gauge invariance of action \eqref{eq:action-bosonic} is realized in the boundary dual as a global $SU(2)$ internal symmetry. The presence of a non-vanishing value of $\phi(z)$ at the boundary is interpreted as a finite chemical potential $\mu$ for the particles charged under such symmetry,  and breaks the $SU(2)$ global symmetry to a residual global $U(1)_{\tau_0}$. The choice of the boundary condition $W_{UV} =0$ corresponds to turning off the source of the dual operator $J_{x}^1$ in the QFT side, whose expectation value is given by $W'_{UV}$. In consequence the quantity $W'_{UV}=\langle J_{x}^1\rangle$ can be considered as the order parameter of the QFT, a non-vanishing value of which $W'_{UV}\neq 0$ signals the spontaneous breaking of the residual global $U(1)_{\tau_0}$ and rotational symmetries.
A solution with non-vanishing $W'_{UV}$ is thus interpreted as the dual of a $p$-wave superconducting phase \cite{GubserP}. On the other hand, the normal state of the system is described by the AdS-Reissner-N\"ordstrom solution (\ref{eq:RN-nonextremal})
with $W'_{UV}=0$.

\begin{figure}[ht]
\begin{center}
\begin{minipage}[t]{\textwidth}
\begin{minipage}{0.5\textwidth}
\center
\subfigure[]{\includegraphics[width=0.85 \textwidth]{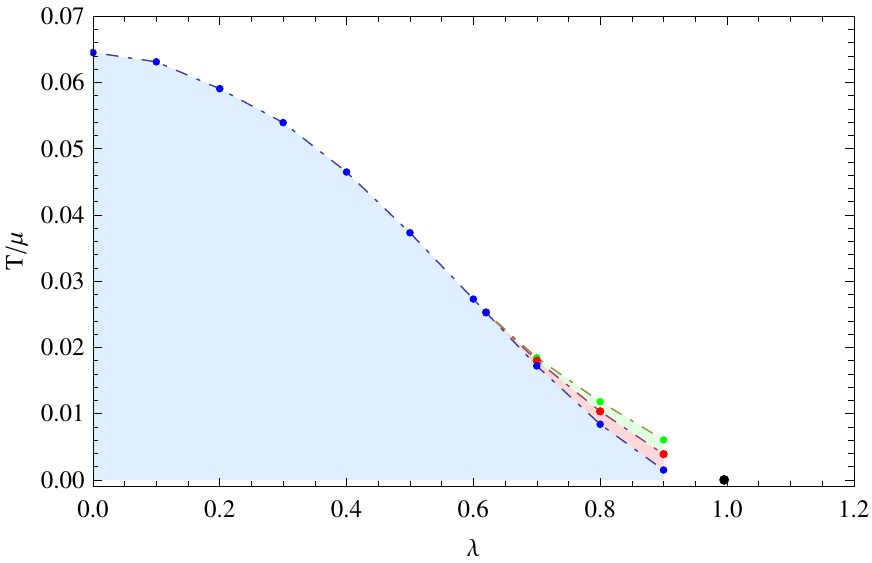}}\vspace{5mm}
\subfigure[]{\includegraphics[width=0.85\textwidth]{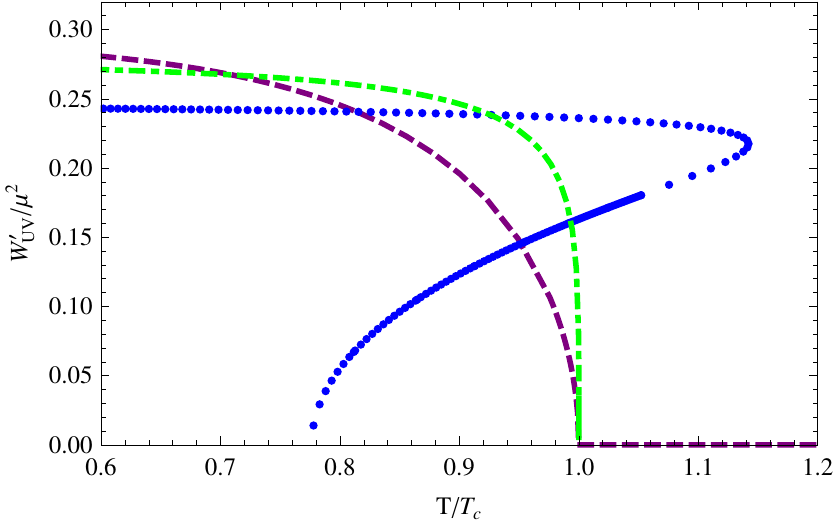}}
\end{minipage}
\renewcommand{\thesubfigure}{(\arabic{subfigure})}
\makeatletter
\makeatother
\setcounter{subfigure}{0}
\ \
\begin{minipage}{0.44\textwidth}
\center \vspace{-0.3cm}
\subfigure[][]
{\includegraphics[width=0.62\textwidth]{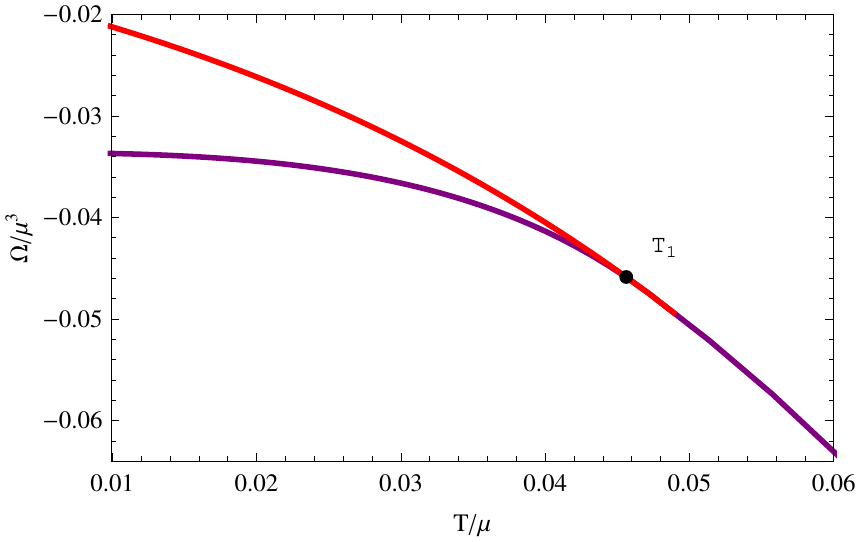}} \vspace{-0.15cm}
\subfigure[][]
{\includegraphics[width=0.62\textwidth]{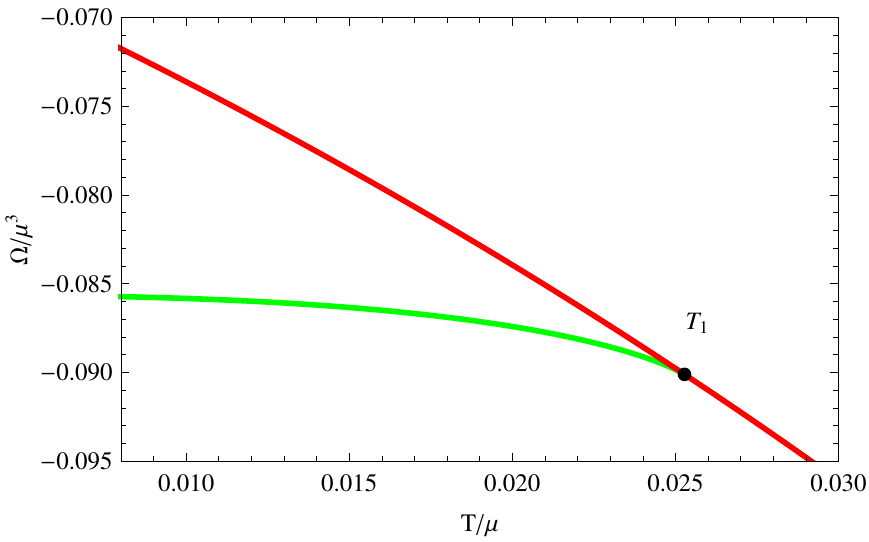}}\vspace{-0.00cm}
\subfigure[][]
{\includegraphics[width=0.62\textwidth]{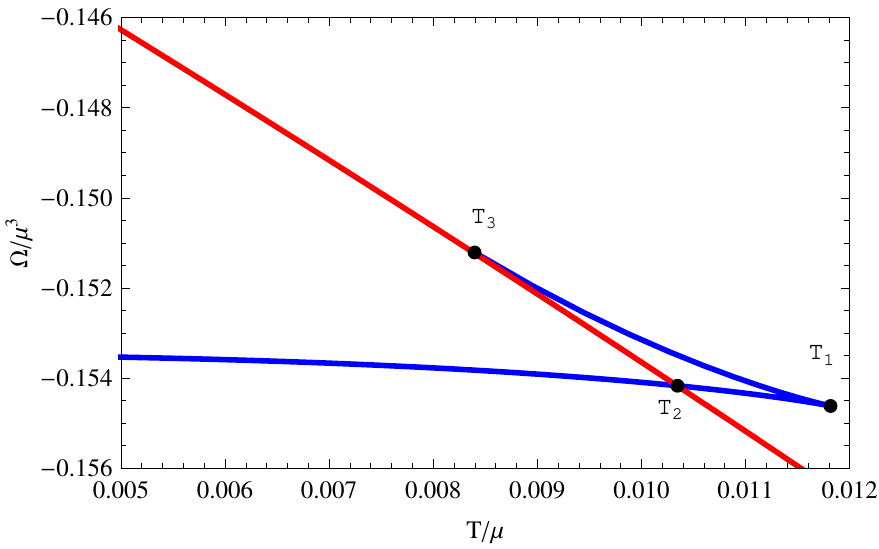}}
\end{minipage}
\end{minipage}
\caption{
(a) Phase diagram of a $p$-wave superconductor in the $\lambda-T$ plane.
The black point indicates the critical coupling $\lambda_*$ at $T=0$  where a quantum phase transition takes place.
(b) The order parameter as a function of the temperature for three different values of $\lambda$: $\lambda=0.40<\lambda_c\,$ (purple), $\lambda=0.62=\lambda_c\,$ (green), $\lambda=0.80>\lambda_c\,$ (blue).
(1) (2) (3) The free energy for the values of $\lambda$ as in (b); the red curves are
the free energies of the system in the normal phase.
\label{fig:pwave-phases}
}
\end{center}
\end{figure}

The Hawking temperature of the bulk solution is interpreted as the temperature of the thermal bath in contact with the boundary theory. Since the AdS-Reissner-N\"ordstrom solution is energetically favoured for temperatures above $T_{c}$, we can say the dual theory is in its normal phase for $T>T_c$, while it goes into a superconducting anisotropic phase for $T<T_{c}$.
The value of $T_c$ and the order of the phase transition depend on the coupling $\lambda$.
In figure \ref{fig:pwave-phases}(a) we display the phase diagram of the $p$-wave superconductor. We can distinguish three regions of the gravitational coupling, namely
\begin{itemize}
\item When $0\leq\lambda\leq\lambda_c\approx 0.62$ the system experiments second order phase transitions from the low temperature, hairy phase to the high temperature, normal phase at some $T_c$ (blue critical line).
\item When $\lambda_c<\lambda\leq\lambda_*\approx 0.995$ the phase transitions become first order (red critical line).
We note the presence of two spinodal curves above and below the critical line that signal the emergence of the two-branched hairy solution ($T_1$) and the disappearance of the unstable branch ($T_2$) respectively.
\item For $\lambda>\lambda_*$ the symmetry broken phase ceases to exist.
\end{itemize}
The order parameter $W'_{UV}$ and the free energies are shown in figures \ref{fig:pwave-phases}(b) and \ref{fig:pwave-phases}(1)(2)(3) respectively, for three different typical couplings $\lambda=0.40<\lambda_c$ (where $T_c = 0.0467607\,\mu$), $\lambda=0.62=\lambda_c$ (where $T_c = 0.0252852\,\mu$) and $\lambda=0.80>\lambda_c$ (where $T_c = 0.0103462\,\mu$).

~

In what follows, after introducing fermions in the bulk, we analyze how their spectral function changes over the whole phase diagram.

\newpage
\section{Fermionic sector: adding fermions to the holographic $p$-wave superconductor}
\label{sec:adding-fermions}
Now let us move on the fermionic sector, introducing fermionic degrees of freedom in the above defined setup. We do so in the gravitational side, and we interpret the numerical results from the QFT point of view.
\subsection{Dirac spinors in the gravitational background}
\label{sec:dirac}
We consider Dirac spinors in the fundamental representation  of $SU(2)$, defined by the generators $\{\tau^0,\tau^1,\tau^2\}$ $=\{\sigma^3/2,\sigma^1/2,\sigma^2/2\}$ where $\sigma_1,\sigma_2,\sigma_3$ are the Pauli matrices. To couple those spinors to gravity, we first introduce a vielbein $\{\omega^A :G=\eta_{AB}\,\omega^A\,\omega^B$\}, together with its dual vector basis $\{e_A : e_A(\omega^B)= \delta_A^B\}$, and a spin connection $\{\omega^A_{\,B}\}$, where the local indices run on $A,B=0,\dots,3$. In the background \eqref{eq:ansatz} they read
\be\label{eq:vielbein}
\begin{array}{llllll}
\omega^0 \equiv \frac{s\sqrt{f}}{z}\;dt\,,
&\quad~\quad&
e_0\equiv \frac{z}{s\sqrt{f}}\;\frac{\partial}{\partial t}\,,
&\quad~\quad&
\omega^0{}_3=-\omega_{03}=+\omega_{30}= -\sqrt{f}\,\left(1-z\,\frac{s'}{s} - \frac{z\,f'}{2\,f}\right)\;\omega^0
\,,
\cr
\omega^1 \equiv \frac{1}{zg}\;dx\,,
&\quad~\quad&
e_1\equiv zg\;\frac{\partial}{\partial x}\,,
&\quad~\quad&
\omega^1{}_3=+\omega_{13}=-\omega_{31}=-\sqrt{f}\,\left(1+z\,\frac{g'}{g}\right)\;\omega^1\,,
\cr
\omega^2 \equiv \frac{g}{z}\;dy\,,
&\quad~\quad&
e_2\equiv \frac{z}{g}\;\frac{\partial}{\partial y}\,,
&\quad~\quad&
\omega^2{}_3=+\omega_{23}=-\omega_{32}=-\sqrt{f}\,\left( 1 - z\,\frac{g'}{g}\right)\;\omega^2\,,
\cr
\omega^3 \equiv \frac{1}{z\sqrt{f}}\;dz\,,
&\quad~\quad&
e_3\equiv z\sqrt{f}\;\frac{\partial}{\partial z}\,.&
&\quad~\quad&~
\end{array}
\ee
Local gamma-matrices obey $\{\Gamma^A,\Gamma^B\} = 2\,\eta^{AB}$, and allow us to define a covariant derivative as usual
\be
\label{eq:covariant-derivative}
\slashed D\Psi \equiv \Gamma^A\,D_A\Psi = \Gamma^A\left( e_A(\Psi) +
\frac{1}{8}\,\omega^{BC}{}_A\,[\Gamma_B,\Gamma_C]\,\Psi - i\, A_A^a\, 
\tau_a
\,\Psi\right)\,,
\ee
and to introduce the projectors on eigenstates of $\Gamma^3$ given by $P_\pm\equiv \frac{1}{2}\,(1\mp\Gamma^3)$ with eigenvalues $\mp 1$.
It is well-known that a boundary term is needed to have a well-defined variational principle \cite{HS}\cite{Henneaux:1998ch}.
If we fix the boundary value of the Dirac fields to be the left (right)-handed part $\Psi_+\equiv P_+\Psi\; (\Psi_-\equiv P_-\Psi)$, the action to be considered is
\be
\label{eq:action-fermionic}
S^{(fer)} = -\int d^4 x\,\sqrt{-G}\;\bar\Psi\,\left(\slashed D - m \right)\,\Psi
\mp\int_{z\rightarrow 0} d^3 x\,\sqrt{-h}\; \bar\Psi_\pm\;\Psi_\mp\,,
\ee
where $\;\bar\Psi\equiv\Psi^{\dagger}\,i\,\Gamma^0\;$ and $\;h=h_{\mu\nu}(x)\,dx^\mu\,dx^\nu\;$
is the induced metric on the boundary.
The resulting equations of motion read
\be
(\slashed D - m)\,\Psi =0\,.
\label{eq:eom-fermionic}
\ee
To solve these equations, we find convenient to work in momentum space and re-scale the spinors as
$
\Psi(x,z) \equiv (-\mbox{det}G\,G^{zz})^{-1/4} \,e^{i\,k_\mu\, x^\mu}\, \psi_k(z)={z^{3/2}}{f^{-\frac{1}{4}}\,{s}^{-\frac12}}\,e^{i\,k_\mu\, x^\mu}\,\psi_k(z)
$
where $(k_\mu)=(-\omega, k_x, k_y)$.
By using the following representation of the Dirac $\Gamma$ matrices
\be
\Gamma^\mu\equiv\left(\begin{array}{cc}0&\gamma^\mu\\ \gamma^\mu&0\end{array}\right)
\,,
\qquad\qquad
\Gamma^{3}\equiv\left(\begin{array}{cc}-1& 0\\ 0&1\end{array}\right)\,,
\label{eq:gamma-matrices}
\ee
where the $2+1$ dimensional $\gamma^\mu$-matrices are:
$\gamma^0=i\,\sigma_2\,,\,\gamma^1=\sigma_1\,,\,\gamma^2=\sigma_3$, we can introduce two-dimensional spinors $\psi^\pm_k(z)$ as
\footnote{
The generators of the Lorentz subgroup in $2+1$ dimensions are,
\be
\frac{i}{4}[\Gamma_0,\Gamma_1]=\frac{i}{2}\,\left(\begin{array}{cc}\sigma_3&0\\0&\sigma_3\end{array}\right)
\,,\quad\quad
\frac{i}{4}[\Gamma_0,\Gamma_2]=\frac{i}{2}\,\left(\begin{array}{cc}-\sigma_1&0\\0&-\sigma_1\end{array}\right)
\,,\quad\quad
\frac{i}{4}[\Gamma_1,\Gamma_2]=\frac{1}{2}\,\left(\begin{array}{cc}\sigma_2&0\\0&\sigma_2\end{array}\right)
\,,\nonumber
\label{eq:loretnz-generators}
\ee
which displays the reducibility of the spinorial representation with respect to $spin(1,2)$, being $i\,\sigma_3 /2$ and $-i\,\sigma_1 /2$ the boosts generators and $\sigma_2 /2$ the generator of rotations in the $(x-y)$-plane.
The left/right colored spinors $\psi^\pm_k(z)$ are thus Dirac spinors of the boundary theory.
}
\be
\psi_k(z) \equiv\left(\begin{array}{l}\psi^+_k(z)\\ \psi^-_k(z) \end{array}\right)
\,.
\label{eq:bispinor}
\ee
The equations of motion (\ref{eq:eom-fermionic}) then reads
\bea\label{ecuacionesgenerales}
+\psi_k^+{}' + \frac{i}{\sqrt{f}}\,\left[\frac{1}{s\sqrt{f}}\left(\omega +\phi\,\tau_0\right)\gamma^{0}-g\left(k_x-W\,\tau_{1}\right)\gamma^{1}-\frac{k_y}{g}\,\gamma^{2}\right]\,\psi_k^-
=-\frac{m}{z\sqrt{f}}\psi_k^+\,,\cr
-\psi_k^-{}'+ \frac{i}{\sqrt{f}}\,\left[\frac{1}{s\sqrt{f}}\left(\omega +\phi\,\tau_0\right)\gamma^{0}-g\left(k_x-W\,\tau_{1}\right)\gamma^{1}-\frac{k_y}{g}\,\gamma^{2}\right]\,\psi_k^+
=-\frac{m}{z\sqrt{f}}\psi_k^-\,.
\eea
Now we can make explicit the color indices by writing
$\psi^\pm_k = (\psi^\pm_k{}^{\alpha i})$ where $\alpha=1,2$ and $i=1,2$ are boundary spinor and color indices respectively.
It will be convenient in what follows to introduce the four-tuples $\vec u_k(z)$ and $\vec v_k(z)$ as
\bea\label{eq:four-tuples}
u_k^1(z)\equiv\psi^+_k{}^{11}(z)\,,\quad\quad
u_k^2(z)\equiv\psi^+_k{}^{21}(z)\,,\quad&&\quad
u_k^3(z)\equiv\psi^+_k{}^{12}(z)\,,\quad\quad
u_k^4(z)\equiv\psi^+_k{}^{22}(z)\,,\cr
v_k^1(z)\equiv \psi^-_k{}^{11}(z)\,,\quad\quad
v_k^2(z)\equiv \psi^-_k{}^{21}(z)\,,\quad&&\quad
v_k^3(z)\equiv \psi^-_k{}^{12}(z)\,,\quad\quad
v_k^4(z)\equiv \psi^-_k{}^{22}(z)\,,
\eea
and the $4\times4$ real matrix
\bea
\label{eq:U}
{\bf U}(z)\equiv \frac{1}{\sqrt{f}}
\left(\begin{array}{cccc}
 -\frac{k_y}{g} &\frac{\left(\omega +\frac{\phi}{2}\right)}{s\,\sqrt{f}}-g\,k_x&0 &g\,W/2\\
-\frac{\left(\omega +\frac{\phi}{2}\right)}{s\,\sqrt{f}}-g\,k_x & \frac{k_y}{g}&g\,W/2&0\\
0&g\,W/2& -\frac{k_y}{g}&\frac{\left(\omega -\frac{\phi}{2}\right)}{s\,\sqrt{f}}-g\,k_x\\
g\,W/2&0&-\frac{\left(\omega -\frac{\phi}{2}\right)}{s\,\sqrt{f}}-g\,k_x&\frac{k_y}{g}
\end{array}\right)\,.
\eea
In terms of this notation, equations (\ref{ecuacionesgenerales}) are compactly written as
\bea
\label{eq:eom-four-tuples}
+\vec u_k'(z) + i\,{\bf U}(z)\,\vec v_k(z) &=& 0
\,,
\cr
-\vec v_k'(z) + i\,{\bf U}(z)\,\vec u_k(z) &=& 0
\,,
\eea
where we put $m=0$ as we concentrate in such particular case in the rest of the paper.
These equations need to be solved for the eight functions $\vec u_k(z), \vec v_k(z)$.
Since they  are first order equations, the solutions are written in terms of eight undetermined constants. These constants are fixed by choosing boundary conditions at the frontiers $z_{UV}=0$  and $z_{IR}$ of the  domain of the variable $z$. We make the choice of fixing the left-handed part $\Psi_+$ at the boundaries, that is $\vec u_k(z)$, the action being as in (\ref{eq:action-fermionic}) with the upper sign. Then, when we impose our boundary conditions at $z_{IR}$, four of those constants are determined. As usual in boundary value problems, the remaining boundary conditions at $z_{UV}=0$ can be satisfied by a non-trivial solution if and only if a certain ``on-shell'' constraint among the free parameters in \eqref{eq:eom-four-tuples} is fulfilled $C(\omega,k_x,k_y)=0$. In what follows, we will be interested in such ``on-shell'' solutions (that following the literature we call ``normal modes''), as well as in ``off-shell'' ones satisfying boundary conditions only at the $z_{IR}$ boundary.

Equations \eqref{eq:eom-four-tuples} can be solved iteratively in the vicinity of any point $z_0$ by writing
\bea
\vec u_k(z) &=& {\cal C}(z;z_0)\,\vec u_k^{(0)}- i\,{\cal S}(z;z_0)\,\vec v_k^{(0)}\quad,
\cr
\vec v_k(z) &=&i\,{\cal S}(z;z_0)\,\vec u_k^{(0)}+{\cal C}(z;z_0)\,\vec v_k^{(0)}\quad,
\label{eq:solution-evolution}
\eea
where
\bea
{\cal C}(z;z_0) &=& {\sf P}\cosh\left(\int^z_{z_0}\,dz'\,{\bf U}(z')\right)\equiv
\frac{1}{2}\left({\sf P}\,e^{\int^z_{z_0}\,dz'\,{\bf U}(z')} +
{\sf P}\,e^{-\int^z_{z_0}\,dz'\,{\bf U}(z')}\right)\quad,\cr
{\cal S}(z;z_0) &=& {\sf P}\sinh\left(\int^z_{z_0}\,dz'\,{\bf U}(z')\right)\equiv
\frac{1}{2}\left({\sf P}\,e^{\int^z_{z_0}\,dz'\,{\bf U}(z')} -
{\sf P}\,e^{-\int^z_{z_0}\,dz'\,{\bf U}(z')}\right)\quad.
\label{eq:evolution}
\eea
Here ${\sf P}$ entails for path ordering. The evolution operators ${\cal C}(z;z_0)$ and ${\cal S}(z;z_0)$ map the values $\vec u_k^{(0)}$ and $\vec v_k^{(0)}$  of $\vec u_k(z)$ and $\vec v_k(z)$ at the point $z_0$ into their values at $z$. We use them to analyze the behavior of the solutions in the IR and UV limits.

\subsubsection{Ingoing boundary conditions for vanishing Beckenstein-Hawking entropy}
\label{sec:fermions-zero-entropy}
When the Beckenstein-Hawking entropy vanish, the boundary conditions must be imposed at $z_{IR}\rightarrow\infty$. The matrix ${\bf U}(z)$ is constant there
\bea
{\bf U}_{IR}\equiv {\bf U}(\infty)=
\left(\begin{array}{cccc}
-\frac{k_y}{g_{IR}}&\frac{\omega}{s_{IR}} - g_{IR}\,k_x&0 &\frac{g_{IR}\,w_{IR}}{2}\\
-\frac{\omega}{s_{IR}} - g_{IR}\,k_x& \frac{k_y}{g_{IR}}&\frac{g_{IR}\,w_{IR}}{2}&0\\
0&\frac{g_{IR}\,w_{IR}}{2}&-\frac{k_y}{g_{IR}}&\frac{\omega}{s_{IR}} - g_{IR}\,k_x\\
\frac{g_{IR}\,w_{IR}}{2}&0&-\frac{\omega}{s_{IR}} - g_{IR}\,k_x&\frac{k_y}{g_{IR}}
\end{array}\right)\,,
\label{eq:U-IR}
\eea
where $(w_{IR}, s_{IR}, g_{IR})$ were introduced in (\ref{eq:boundary-IR-zero-entropy}).
The eigenvalues of ${\bf U}_{IR}$  are given by $\left(\kappa_+, -\kappa_+, \kappa_-, -\kappa_-\right)$ satisfying
\be\label{eq:kappa+-}
\kappa_{\pm}{}^2 =
g^{2}_{IR}\left(k_x\pm\frac{w_{IR}}{2}\right)^{2}+\frac{k_y^{2}}{g^{2}_{IR}}
-\frac{\omega^2}{s^{2}_{IR}}\,.
\ee
The matrix ${\bf U}_{IR}{}^2$ is symmetric (although ${\bf U}_{IR}$ it is not) and then it is diagonalizable by an orthogonal matrix $P$ such that $P^t\,{\bf U}_{IR}{}^2\,P ={\mbox{Diag}}( \kappa_+{}^2, \kappa_+{}^2, \kappa_-{}^2, \kappa_-{}^2)$
and
\be
P\equiv \frac{1}{\sqrt{2}}\left(\begin{array}{cc}1_{2\times 2} &1_{2\times 2}\\-1_{2\times 2} & 1_{2\times 2}\end{array}\right)\,.
\label{eq:matriz-S}
\ee
Let us take $z_0=z_{ir}\gg 1$, and let us consider the region $z>z_{ir}\gg 1$.
There the evolution operators read
\ba&&
{\cal C}(z;z_{ir})\simeq P\,\left(\begin{array}{cc}
\cosh(\kappa_+(z-z_{ir}))\,1_{2\times 2}&0_{2\times 2}\\0_{2\times 2}&\cosh(\kappa_-(z-z_{ir}))\,1_{2\times 2}\\\end{array}\right)\,P^{t}\,,
\cr&&
{\cal S}(z;z_{ir})\simeq P
\left(\begin{array}{cc}
\sinh(\kappa_+(z-z_{ir}))\,1_{2\times 2}&0_{2\times 2}\\0_{2\times 2}&\sinh(\kappa_-(z-z_{ir}))\,1_{2\times 2}\\\end{array}\right) D\, P^{t}\,{\bf U}_{IR} \,,
\label{eq:evolution-IR}
\ea
where $D= \mbox{Diag}(1/\kappa_+,1/\kappa_+,1/\kappa_-,1/\kappa_-)$.
Now equations \eqref{eq:solution-evolution} implies that in the large $z$ limit we have
\bea
{\vec u}_k(z) &\simeq& \frac{1}{2} P\left(\begin{array}{cc}
e^{\kappa_+(z-z_{ir})}\,1_{2\times 2}&0_{2\times 2}\\0_{2\times 2} & e^{\kappa_-(z-z_{ir})}\,1_{2\times 2}\\
\end{array}\right)\; \left(P^t\,\vec u_k^{(ir)}-i\,\tilde D\,P^{t}\,{\bf U}_{IR}\,\vec v_k^{(ir)}\right)+\cr
&&+\, \frac{1}{2} P\left(\begin{array}{cc}
e^{-\kappa_+(z-z_{ir})}\,1_{2\times 2}&0_{2\times 2}\\0_{2\times 2} & e^{-\kappa_-(z-z_{ir})}\,1_{2\times 2}\\
\end{array}\right)\; \left(P^t\,\vec u_k^{(ir)}+i\,\tilde D\,P^{t}\,{\bf U}_{IR}\,\vec v_k^{(ir)}\right)\,,\cr
&&~ \cr
{\vec v}_k(z) &\simeq& \frac{1}{2} P\left(\begin{array}{cc}
e^{\kappa_+(z-z_{ir})}\,1_{2\times 2}&0_{2\times 2}\\0_{2\times 2} & e^{\kappa_-(z-z_{ir})}\,1_{2\times 2}\\
\end{array}\right)\; \left(P^t\,\vec v_k^{(ir)}+i\,\tilde D\,P^{t}\,{\bf U}_{IR}\,\vec u_k^{(ir)}\right)+\cr
&&+\, \frac{1}{2} P\left(\begin{array}{cc}
e^{-\kappa_+(z-z_{ir})}\,1_{2\times 2}&0_{2\times 2}\\0_{2\times 2} & e^{-\kappa_-(z-z_{ir})}\,1_{2\times 2}\\
\end{array}\right)\; \left(P^t\,\vec v_k^{(ir)} - i\,\tilde D\,P^{t}\,{\bf U}_{IR}\,\vec u_k^{(ir)}\right)\,.
\label{eq:comportamiento-IR}
\eea
Let us first consider $\kappa_\pm{}^2>0$, and then $\kappa_\pm\equiv +\sqrt{\kappa_\pm{}^2}$.
The condition that the solution to be well-behaved implies that the coefficients of the positive exponentials must vanish, yielding the solution
\bea
{\vec u}_k(z) &\simeq& P\left(\begin{array}{cc}
e^{-\kappa_+(z-z_{ir})}\,1_{2\times 2}&0_{2\times 2}\\0_{2\times 2} & e^{-\kappa_-(z-z_{ir})}\,1_{2\times 2}\\
\end{array}\right)\; P^t\,\vec u_k^{(ir)}\,,\cr
&&~\cr
{\vec v}_k(z) &\simeq& P\left(\begin{array}{cc}
e^{-\kappa_+(z-z_{ir})}\,1_{2\times 2}&0_{2\times 2}\\0_{2\times 2} & e^{-\kappa_-(z-z_{ir})}\,1_{2\times 2}\\
\end{array}\right)\;P^t\,\vec v_k^{(ir)}\,,
\label{eq:comportamiento-IR_ibc S=0}
\eea
where the initial values of the fields satisfy the linear relation
\be\label{eq:linearS=0}
\vec u_k^{(ir)}=i\, P\,\tilde D\,P^{t}\,{\bf U}_{IR}\,\vec v_k^{(ir)}\,.
\ee
Similar result is obtained when $\kappa_+{}^2<0$ and/or $\kappa_-{}^2<0$ under the imposition of ingoing boundary conditions, with the replacement $\kappa_\pm\rightarrow -i\, sign(\omega)\, \sqrt{-\kappa_\pm{}^2}$ in (\ref{eq:comportamiento-IR_ibc S=0}). Regarding the case $\kappa_\pm{}^2=0$, that will be of particular importance later, we get
\ba&&
{\cal C}(z;z_{ir})\simeq 1\,,
\cr&&
{\cal S}(z;z_{ir})\simeq 0 \,,
\label{eq:evolution-IR-zero}
\ea
implying the behaviors
\bea
{\vec u}_k(z) &\simeq& \vec u_k^{(ir)}\,,\cr
{\vec v}_k(z) &\simeq& \vec v_k^{(ir)}\,.
\label{eq:comportamiento-IR_ibc S=0-zero}
\eea
Notice that these are regular at the horizon witout imposing any additional relation between
$\vec u_k^{(ir)}$ and $\vec u_k^{(ir)}$.
\subsubsection{Ingoing boundary conditions for non-vanishing Beckenstein-Hawking entropy}
\label{sec:fermions-non-zero-entropy}
When the Beckenstein-Hawking entropy does not vanish, the horizon sits at $z_h =1$,
and the matrix ${\bf U}(z)$ is not a constant matrix there but presents a simple pole
\be
{\bf U}(z)\stackrel{z\rightarrow 1^-}{\longrightarrow}
\frac{-i\,\omega}{4\,\pi\,T\,(1-z)}\;{\bf C}\,,\qquad\qquad
{\bf C}\equiv\left(\begin{array}{cc}i\,\gamma^0&0_{2\times 2}\\0_{2\times 2}&i\,\gamma^0\end{array}\right)\,,
\label{eq:U-T}
\ee
where $T$ is the Hawking temperature introduced in (\ref{eq:BHtemp}). Notice that ${\bf C}$ is pure imaginary, hermitian and satisfies ${\bf C}^2=1$. Let us take now $|1-z_{ir}|\ll 1$, and let us look at the near horizon region $z_{ir}<z\leq 1$. The evolution operators (\ref{eq:evolution}) at leading order then read
\be
{\cal C}(z; z_{ir})\simeq\cos\left(\frac{\omega}{4\,\pi\,T} \ln\frac{1-z}{1-z_{ir}}\right){\bf 1}\,,\qquad\qquad
{\cal S}(z; z_{ir})\simeq i\,\sin\left(\frac{\omega}{4\,\pi\,T}\ln\frac{1-z}{1-z_{ir}}\right){\bf C}\,,
\label{eq:evolutionT}
\ee
From (\ref{eq:solution-evolution}) they yield the solution
\bea
\vec u_k(z)&\simeq&
\cos\left(\frac{\omega}{4\,\pi\,T}\ln\frac{1-z}{1-z_{ir}}\right)\,\vec u_k^{(ir)}+
\sin\left(\frac{\omega}{4\,\pi\,T}\ln\frac{1-z}{1-z_{ir}}\right)\,{\bf C}\,\vec v_k^{(ir)}\,,\cr
\vec v_k(z)&\simeq&
-\sin\left(\frac{\omega}{4\,\pi\,T}\ln\frac{1-z}{1-z_{ir}}\right)\,{\bf C}\,\vec u_k^{(ir)}
+\cos\left(\frac{\omega}{4\,\pi\,T}\ln\frac{1-z}{1-z_{ir}}\right)\,\vec v_k^{(ir)}\,.
\label{eq:comportamientoT}
\eea
Ingoing boundary conditions at the horizon fix again a linear relation between $\vec u_k^{(ir)}$ and $\vec v_k^{(ir)}$, as
\be\label{eq:linearSno0}
\vec u_k^{(ir)}= i\,{\bf C}\,\vec v_k^{(ir)}\,.
\ee
After imposing this condition the near horizon solution reads,
\bea
\vec u_k(z)&\simeq& e^{-i\,\frac{\omega}{4\,\pi\,T}\ln\frac{1-z}{1-z_{ir}}}\,\vec u_k^{(ir)}\,,\cr
\vec v_k(z)&\simeq& e^{-i\,\frac{\omega}{4\,\pi\,T}\ln\frac{1-z}{1-z_{ir}}}\,\vec v_k^{(ir)}\,.
\label{eq:comportamiento-IR_ibc Sno0}
\eea
\subsubsection{The ultraviolet behavior}
\label{sec:UV sn}
Let us take now $z_0 = z_{UV}=0$ in equations \eqref{eq:solution-evolution}, and consider the region close to the AdS boundary
$z\rightarrow 0$. There the matrix ${\bf U}(z)$ goes to a constant
\bea
{\bf U}_{UV}\equiv {\bf U}(0)=
\left(\begin{array}{cccc}
 -k_y&\omega +\frac{\mu}{2}-k_x&0 &0\\
-\omega-\frac{\mu}{2} -k_x& k_y&0&0\\
0&0& -k_y&\omega -\frac{\mu}{2}-k_x\\
0&0&-\omega+\frac{\mu}{2}-k_x&k_y
\end{array}\right)\,,
\label{eq:U-UV}
\eea
and the evolution operators \eqref{eq:evolution} at leading order are
\be
{\cal C}(z;0)\simeq 1\,,\qquad\qquad {\cal S}(z;0)\simeq{\bf U}_{UV}\;z\,.
\label{eq:evolution-boundary}
\ee
Plugging back into \eqref{eq:solution-evolution}, the leading behavior results
\bea
\vec u_k(z) &\simeq& \vec u_k^{(UV)}- i\,z\,{\bf U}_{UV}\,\vec v_k^{(UV)}\,,\cr
\vec v_k(z) &\simeq& \vec v_k^{(UV)} + i\,z\,{\bf U}_{UV}\,\vec u_k^{(UV)}\,.
\label{eq:solution-evolutionUV}
\eea
\subsubsection{Normal modes and the Dirac cones}
\label{sec:normal-modes}
Normal modes can be defined generically  as the ``most normalizable" regular solutions with real energy. In the fermionic massless case we are considering, the behavior close to the AdS boundary is given by (\ref{eq:solution-evolutionUV}). The natural way to define normal modes should be to search for solutions where the leading terms at the boundary vanish. However, according to the boundary terms included in our action \eqref{eq:action-fermionic}, the best thing we can do is to define fermionic normal modes as regular solutions with $\vec u_k^{(UV)}\equiv\vec u_k(z_{UV})=\vec 0$. Due to the smoothness/ingoing boundary conditions on the horizon, the remaining components $\vec v_k^{(UV)}\equiv\vec v_k(z_{UV})$ are then completely determined, and no condition can be imposed on them. In fact, let us take $z_0=z_{ir}$ and $z=z_{UV}=0$ in (\ref{eq:solution-evolution})
\bea
\vec u_k^{(UV)} &=& {\cal C}(z_{UV};z_{ir})\,\vec u_k^{(ir)}- i\,{\cal S}(z_{UV};z_{ir})\,\vec v_k^{(ir)}\,,
\cr
\vec v_k^{(UV)} &=& i\,{\cal S}(z_{UV};z_{ir})\,\vec u_k^{(ir)}+{\cal C}(z_{UV};z_{ir})\,\vec v_k^{(ir)}\,.
\label{eq:solution-IR-UV1}
\eea
But as we saw in subsections \ref{sec:fermions-zero-entropy} and \ref{sec:fermions-non-zero-entropy} a linear relation $\vec v_k^{(ir)} = -i\,L\,\vec u_k^{(ir)}$ holds where, according to (\ref{eq:linearS=0}) or (\ref{eq:linearSno0}), the unimodular matrix $L$ is given by $L= U_{IR}\,P\,\tilde D\,P^t$ or $L= \bf C$ respectively\footnote{
The normal modes are strictly defined at $T=0$, so we will restrict to this case.
}.
Therefore (\ref{eq:solution-IR-UV1}) can be written as
\bea
\vec u_k^{(UV)} &=& \left({\cal C}(z_{UV};z_{ir})- {\cal S}(z_{UV};z_{ir})\,L\right)\,\vec u_k^{(ir)}\quad,
\cr
\vec v_k^{(UV)} &=& i\,\left({\cal S}(z_{UV};z_{ir})-{\cal C}(z_{UV};z_{ir})\,L\right)\vec u_k^{(ir)}\quad,
\label{eq:solution-IR-UV2}
\eea
which shows that for general momenta $\vec v_k^{(UV)}$ is determined once $\vec u_k^{(UV)}$ is given. However, for normal modes we fix the particular boundary condition $\vec u_k^{(UV)}=0$, and so there can exist a non trivial solution iff
\be
C(\omega, k_x,k_y)\equiv\det\left({\cal C}(z_{UV};z_{ir})- {\cal S}(z_{UV};z_{ir})\,L\right)=0\,.
\label{eq:dirac-cones-exact}
\ee
This equation defines a constraint in momentum space for any value of the frequency, where the normal modes live. Now, let us give some heuristic arguments to analyze some limits of this equation:
\begin{itemize}
\item
In first term let us consider highly energetic particles. Since they probe the near boundary region being essentially blind to the IR region, we can use \eqref{eq:evolution-boundary} as an approximation for their evolution operators, i.e. ${\cal C}(0; z_{ir})\simeq {\cal C}(z_{ir}; 0)\simeq 1\;,\; {\cal S}(0; z_{ir})\simeq -{\cal S}(z_{ir}; 0)\simeq -{\bf U}_{UV}\,z_{ir}$.  In such case, equation (\ref{eq:dirac-cones-exact}) reads
\be
C(\omega, k_x,k_y)\simeq\det\left(1 + z_{ir}\,{\bf U}_{UV}\;L\right)\simeq z_{ir}{}^4\;\det\left({\bf U}_{UV}\right)=0
\label{eq:dirac-cones-UV}
\ee
Therefore in this limit the normal modes have their momentum and frequency such that the eigenvalues of ${\bf U}_{UV}$, given by $\left(\lambda_+, -\lambda_+, \lambda_-, -\lambda_-\right)$, with
\be\label{eq:lambda+-}
\lambda_\pm{}^2 \equiv  k_x{}^2+k_y{}^2 - \left(\omega\pm\frac{\mu}{2}\right)^2\,,
\ee
vanish.
The equations $\lambda_\pm{}^2 = 0$ can then be taken as an approximation of the constraint $C(\omega, k_x,k_y)=0$ for large $|\omega|$. They define the two UV Dirac cones $C^{UV}_\pm$ \cite{Gubser:2010dm}, whose apexes are located at $(\omega, k_x, k_y)= (\mp{\mu}/{2}, 0, 0)$, and that satisfy $\;C^{UV}_+\bigcap C^{UV}_- = \{(0,k_x,k_y) : k_x^2+k_y^2 = {\mu^2}/{4}\}$. Notice that they present rotational symmetry in the momentum plane, indicating that such symmetry is preserved in the UV.
\item
On the other hand, if we consider low energy modes, they probe the large $z$ region being essentially blind to the UV part of the geometry. Then for $\kappa_\pm^2\neq0$ we can use (\ref{eq:evolution-IR}) as an approximation for their evolution operators. The constraint (\ref{eq:dirac-cones-exact}) this time becomes
\be
C(\omega, k_x,k_y)\simeq
e^{(\kappa_++\kappa_-)\,z_{ir}}
=0\,
\ee
which cannot be satisfied for any choice of momentum and frequency. Then we are left with the case $\kappa_\pm^2=0$, where
\be
\kappa_{\pm}{}^2 =
g^{2}_{IR}\left(k_x\pm\frac{w_{IR}}{2}\right)^{2}+\frac{k_y^{2}}{g^{2}_{IR}}
-\frac{\omega^2}{s^{2}_{IR}}\,.
\ee
So the constraint $C(\omega, k_x,k_y)=0$ can be approximated as $\kappa_{\pm}{}^2=0$.
This defines the two IR Dirac cones $C^{IR}_\pm$ \cite{Gubser:2010dm}.
Their apexes are located at $(\omega,k_x,k_y)= (0, \mp{w_{IR}}/{2}, 0)$, and they satisfy
$C^{IR}_+\bigcap C^{IR}_- = \{(\omega, 0, k_y) :\omega^2 =s_{IR}^2 (g_{IR}^2 w_{IR}^2/4 +k_y^2/g_{IR}^2)\}$.
Notice that rotational symmetry in the momentum plane is broken in the IR.
\end{itemize}

\subsection{Fermionic operators in the QFT}
\label{sec:fermions-in-the-QFT}
From the AdS/CFT point of view, the above defined spinor field in the bulk has a spinorial fermionic dual operator ${\cal O}(x)$ in the boundary QFT, which transforms in the doublet representation of $SU(2)$. The two point correlation functions of such operator can be computed by deriving the on-shell action with respect to a source identified roughly with the boundary value of the spinor field. In this subsection we present the computation of the retarded fermionic correlator closely following references \cite{Gubser:2010dm}.

When we evaluate the action (\ref{eq:action-fermionic}) on a solution of the equations of motion only the boundary term contributes. In terms of the vector notation (\ref{eq:four-tuples}) it takes the form
\be
S^{(fer)}_{on-shell}=-\int\frac{d^3 k}{(2\pi)^3}\;\vec u_k^{(UV)}{}^\dagger\;{\bf C}\;\vec v_k^{(UV)}
\label{eq:linear-boundary-horizon}
\ee
However, as noted before, a relation exist for general momenta between the boundary value of the fields; from (\ref{eq:solution-IR-UV2}) we get
\bea
\label{eq:M}
\vec v_k^{(UV)} &=& {\bf M}_k\; \vec u_k^{(UV)} \cr
{\bf M}_k &\equiv& i\,\left({\cal S}(z_{UV};z_{ir})-{\cal C}(z_{UV};z_{ir})\,L\right)\,
\left({\cal C}(z_{UV};z_{ir})- {\cal S}(z_{UV};z_{ir})\,L\right)^{-1}
\eea
Then the on-shell action is
\be\label{eq:action-on-shell}
S^{(fer)}_{on-shell} = -\int \frac{d^3 k}{(2\pi)^3}\;\vec u_k^{(UV)}{}^\dagger\;
{\bf C}\;{\bf M}_k\;\vec u_k^{(UV)}\,.
\ee
In our four-tuple notation, the sources $\vec u_k^{(UV)}$ couple to a four-tuple fermionic operator $\vec{\cal O}(x)= ({\cal O}^I(x), I=1,\dots,4)$ in the boundary theory action, by a term of the form $\delta S=\vec u_k^{UV}{}^\dagger\,{\bf C}\,\vec{\cal O} + h.c.$.
The AdS/CFT correspondence establishes that the connected retarded Green function of them in momentum space can be computed by means of the recipe
\be
\label{eq:green-function}
-i\,\langle\vec{\cal\tilde O}(k)\;\vec{\cal\tilde O}^\dagger(0)\rangle_c|_{ret} \equiv (2\,\pi)^3\,\delta^3( k)\;{\bf G}_R(k)
= {\bf C}\,\frac{\,\,\,\delta^2 S^{(fer)}_{on-shell}}{\delta{\vec u}_k^{(UV)}{}^\dagger\;\delta{\vec u}_0^{(UV)}}\;{\bf C}\,.
\ee
From here, using equation (\ref{eq:action-on-shell}), we can read the correlation function
\be
\label{eq:GMC}
{\bf G}_R(k)= {\bf M}_k\,{\bf C}\,.
\ee

From the point of view of the fermionic theory, the relevant observable to analyze the fermionic dynamics is the spectral function $\rho(k)$, defined as
\be
\label{eq:rho}
\rho(k)\equiv-\mbox{Im\,Tr}\,{\bf G}_R(k)= -\mbox{Im\,Tr}\left({\bf M}_k\,{\bf C}\right)\,.
\ee
It is a non-negative quantity, a fact that from the holographic point of view was shown in \cite{Gubser:2010dm} in the $T=0$ case; we review and extend the analysis for finite $T$ in the appendix A. For translational invariant weakly coupled theories, a K\"ahllen-Lehmann representation for $G_R(k)$ implies that $\rho(k)$ generically presents  delta-like singularities (peaks) as a function of the frequency, on the quasiparticle states of definite dispersion relation, see for example \cite{Landau9}.

Of particular relevance is the spectral function valued at zero frequency $\rho(0,k_x,k_y)$, whose peaks will be labeled $(k^*_x,k^*_y)$. At vanishing temperature, these peaks correspond to the Fermi momenta $(k^F_x,k^F_y)$, and their locus in momentum space determines the Fermi surface. Notice that the normal modes at $\omega=0$ sit on such surface, since the matrix ${\bf M}_k$ is singular for them.

It is clear that is of most importance to determine where the spectral function is non-zero.
From its definition (\ref{eq:rho}) the equation $\,\rho(k) = 0$ is satisfied when $\mbox{Tr}\,{\bf G}_R(k)$ is real. In particular, since ${\bf C}$ is pure imaginary, this happens when ${\bf M}_k$ is pure imaginary. At vanishing entropy, when both $\kappa_\pm{}^2>0$ the boundary conditions in the IR are compatible with a real $\vec u_k^{(ir)}$ and a pure imaginary $\vec v_k^{(ir)}$ since $L$ is real. Now,  being the matrix ${\bf U}(z)$ real, the evolution operators ${\cal C}(z;z_0)$ and ${\cal S}(z;z_0)$ are also real, implying a real $\vec u_k^{(UV)}$ and a pure imaginary $\vec v_k^{(UV)}$. We conclude from (\ref{eq:M}) that the matrix ${\bf M}_k$ is pure imaginary. Then equation (\ref{eq:rho}) implies that $\rho =0$ if both $\kappa_\pm{}^2>0$ and therefore it is non zero when some of the $\kappa_\pm{}^2<0$, that is, inside the IR Dirac cones. Furthermore, when we are at the special momenta that define the normal modes, ${\bf G}_R(k)$ clearly diverges. It can be shown that $\rho$ presents delta-like peaks there \cite{Gubser:2010dm}. On the other hand, at non-vanishing entropy the behavior in the infrared is not compatible with real $\vec u_k^{(UV)}$ and pure imaginary $\vec v_k^{(UV)}$ (because $L$ is pure imaginary) and then in principle $\rho(k)\neq 0$.
\section{Numerical Results}
\label{sec:results}
We perform an extensive numerical study of parameter space, to analyze how the gravitational coupling affects the spectral function as the temperature is varied. The results are contained in the figures displayed in this section. We plotted the spectral function as a function  of $(k_x,k_y)$ for different values of $T$ at fixed $\omega$ and $\lambda$, and as a function of $\omega$ for fixed values of $(k_x,k_y)$, for different values of $\lambda$ and $T$.

\begin{figure}[ht]
\centering
\subfigure[$\,{(k_x,k_y)}= ({k_*}\approx (0.27,0)\mu$]
{\includegraphics[width=53mm]{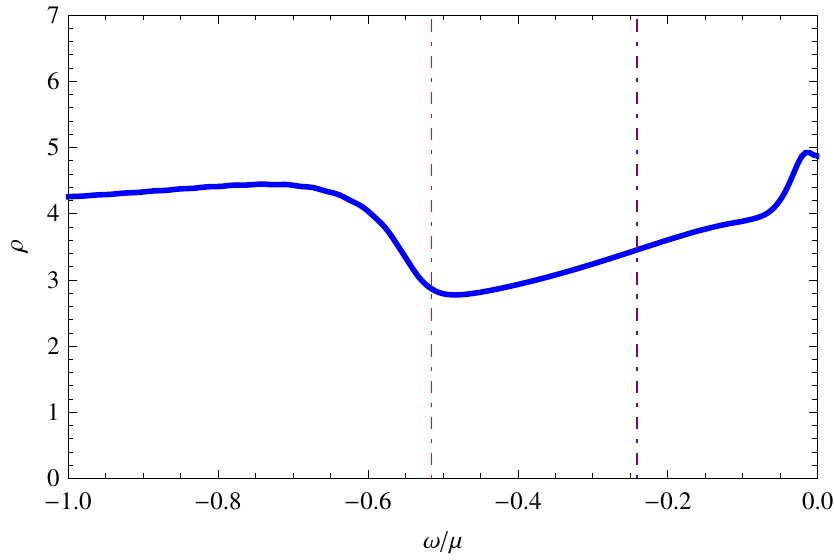}}
\subfigure[$\,(k_x,k_y) =(0.2,\,0.1)\mu $]
{\includegraphics[width=53mm]{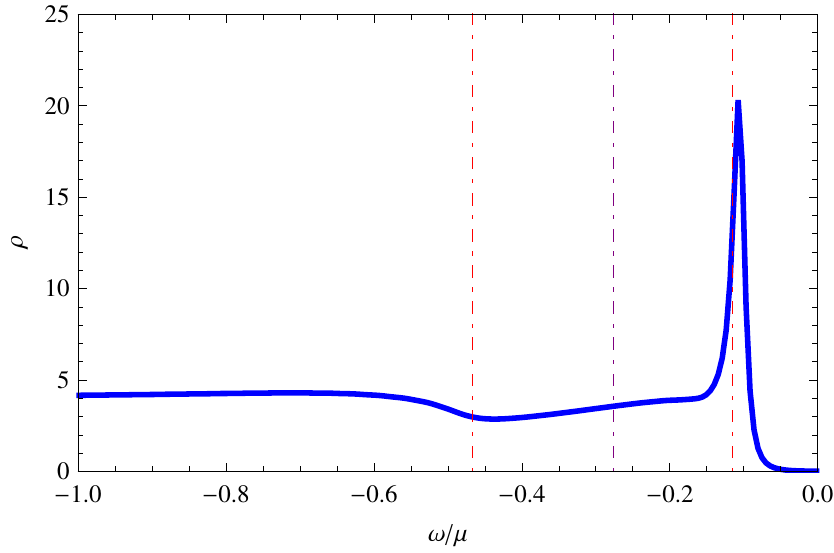}}
\subfigure[$\,{(k_x,k_y)} =(0.2,\,0.4)\mu $]
{\includegraphics[width=53mm]{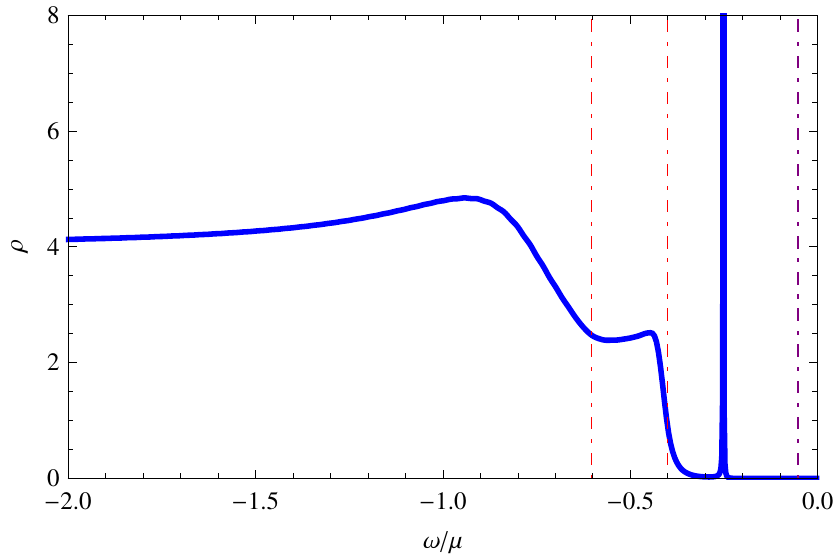}}
\caption{
The spectral function as function of $\omega$ for different $(k_x,k_y)$, at low temperature $T=0.20\, T_{c}$ and $\lambda={1}/{(8\sqrt{2})}$.
\label{fig:gu}
}
\end{figure}

In figure \ref{fig:gu} we present the plots of  the spectral function as a function of $\omega$ for fixed $(k_x,k_y)$ and low $\lambda<\lambda_c$, and a low temperature $T<T_c$. It is seen that the intersections with the IR light-cones are smoothed at non-zero temperature with respect to the $T=0$ case, as it can be seen by comparing  with figure $8$ of reference \cite{Gubser:2010dm}. In the small $\omega$ region, the spectral function oscillates giving rise to a single peak and then it goes to a constant non-vanishing value for $\omega=0$.

\begin{figure}[H]
\centering
\subfigure[ $(k_x,k_y)=(k_*, 0)$]
{\includegraphics[scale=0.6]{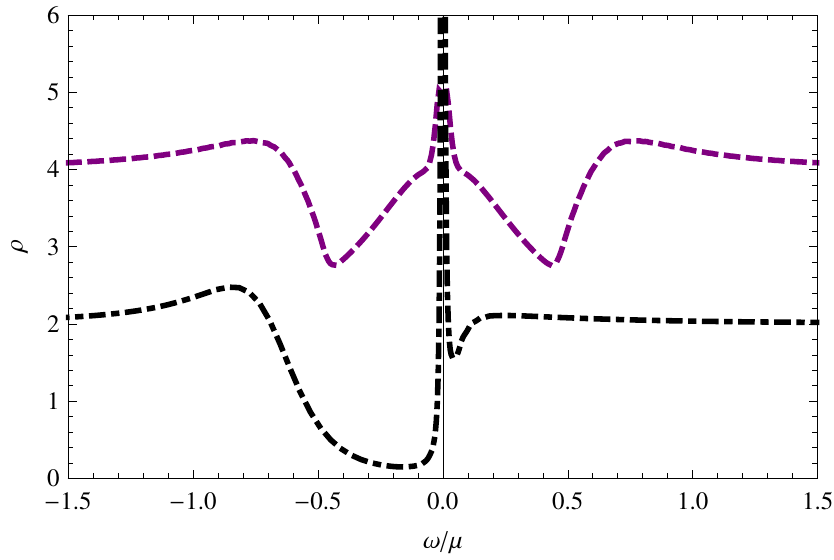}\label{t21} }
\subfigure[$(k_x,k_y)=(0.2,0.1)\mu $]
{\includegraphics[scale=0.6]{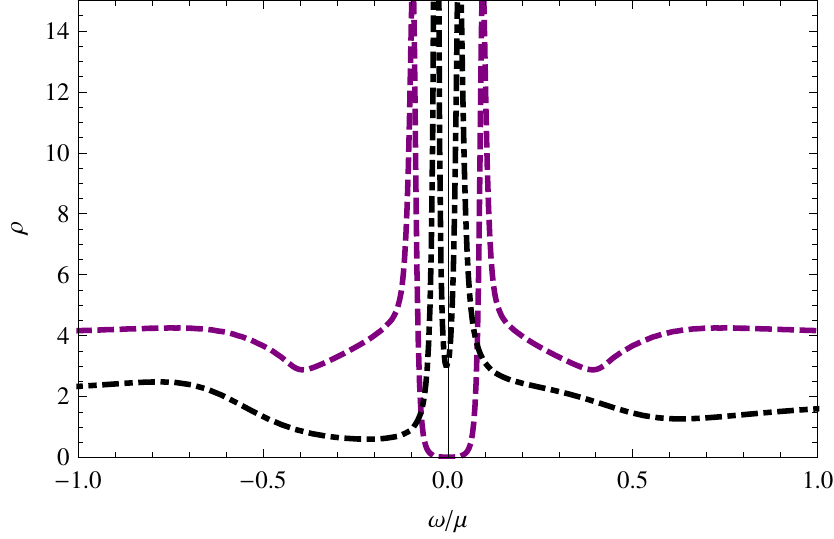}\label{t32} }
\subfigure[$(k_x,k_y)=(0.2,0.4)\mu $]
{\includegraphics[scale=0.5]{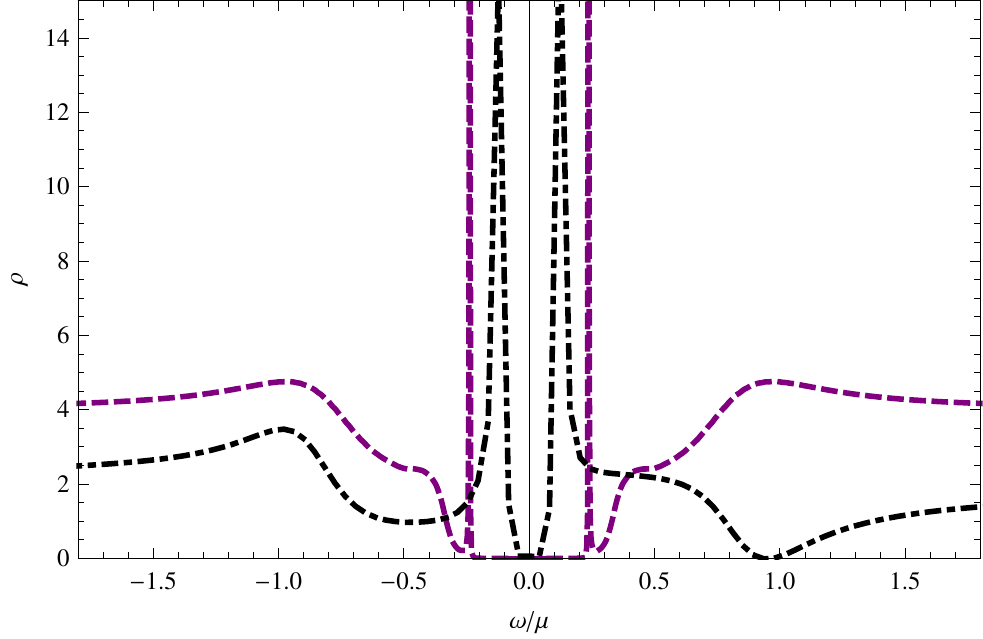}\label{t43} }\\
\subfigure[$(k_x,k_y)=(k_*, 0)$]
{\includegraphics[scale=0.5]{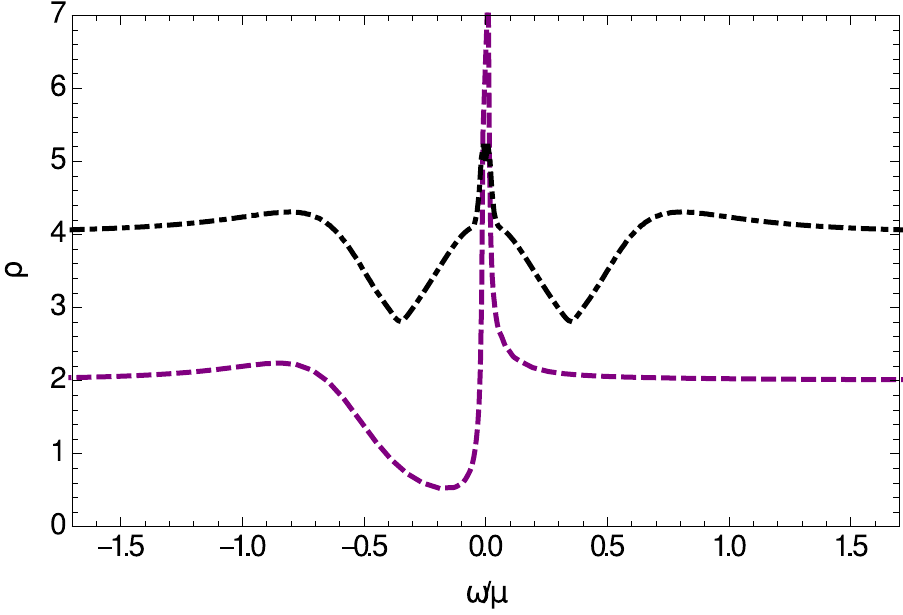}\label{t54}}
\subfigure[$(k_x,k_y) =(0.2,0.1)\mu $]
{\includegraphics[scale=0.6]{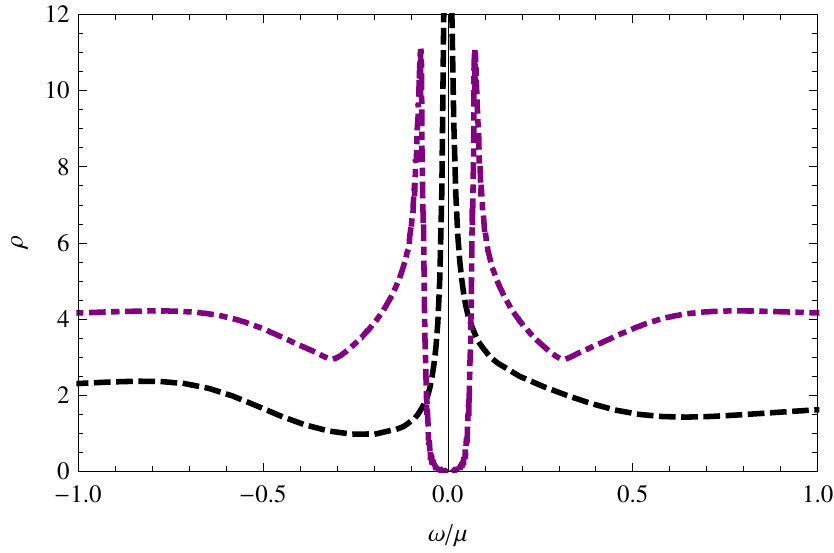}\label{t65} }
\subfigure[$(k_x,k_y) =(0.2,0.4)\mu $]
{\includegraphics[scale=0.6]{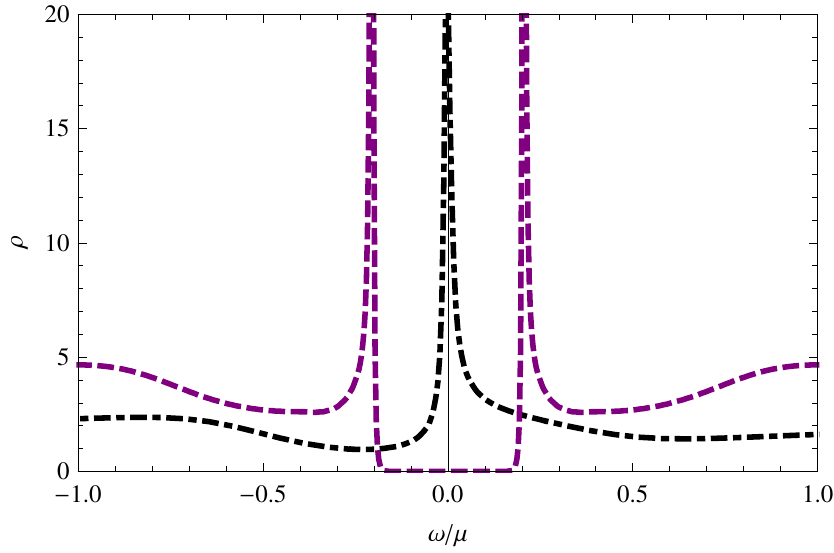}\label{t76} }\\
\subfigure[ $(k_x,k_y)=(k_*, 0)$]
{\includegraphics[scale=0.5]{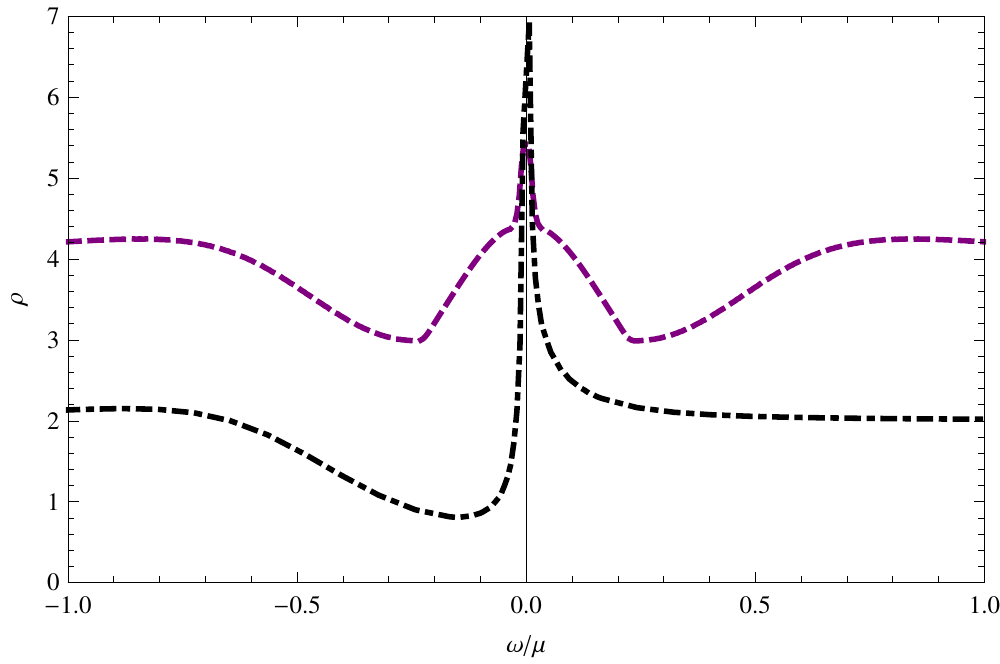}\label{t87} }
\subfigure[$(k_x,k_y) =(0.2,0.1)\mu $]
{\includegraphics[scale=0.6]{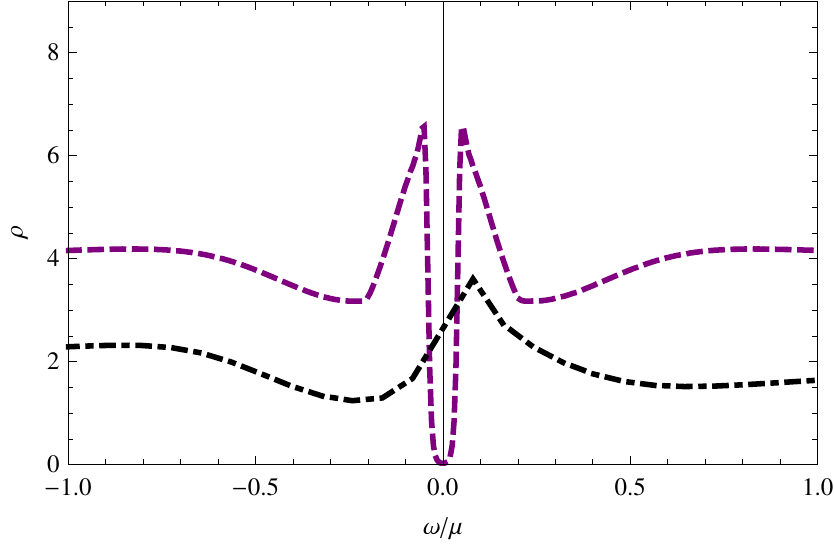}\label{t98} }
\subfigure[$(k_x,k_y) =(0.2,0.4)\mu $]
{\includegraphics[scale=0.6]{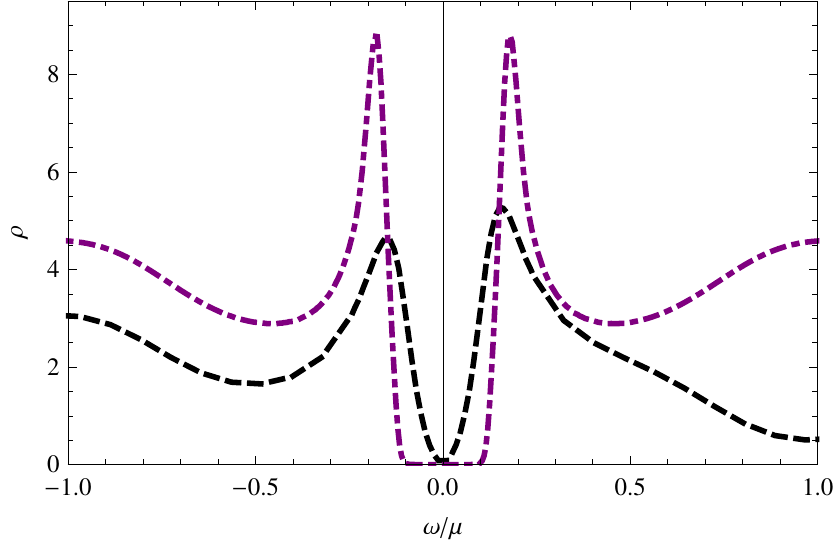}}
\caption{The spectral function as function of $\omega$ in the normal (black) and superconducting (purple) phases,
at fixed $(k_x,k_y)$, for three different temperatures and couplings:
(a) (b) (c) $T=0.20\,T_c\,,\,\lambda=0.40<\lambda_c\,$;
(d) (e) (f) $T=0.28\,T_c\,,\,\lambda=0.62=\lambda_c\,$;
(g) (h) (i) $T=0.40\,T_c\,,\,\lambda=0.80>\lambda_c\,$.
\label{fig:fase-normal-vs-S-CTfin}
}
\end{figure}

In figure \ref{fig:fase-normal-vs-S-CTfin} we give the spectral function in the normal and superconducting phase at $T<T_c$ for different momenta. Is is observed the absence of parity symmetry in $\omega$ in the normal phase in agreement with \cite{Liu:2009dm}, and the presence of such symmetry in the superconducting phase, result observed for the $T=0$ case in \cite{Gubser:2010dm}.

\begin{figure}[H]
\centering
\subfigure[$T=T_{c}$]{\includegraphics[width=0.42\textwidth]{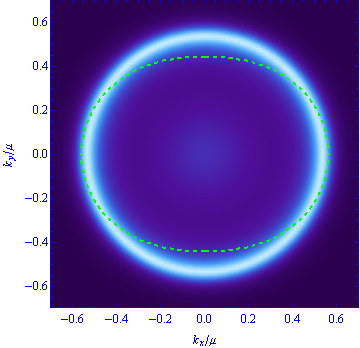}} \hfill
\subfigure[$T=0.85\,T_c$]{\includegraphics[width=0.42\textwidth]{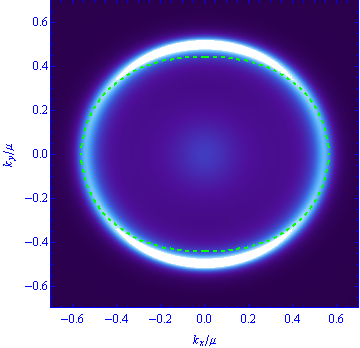}} \vfill
\subfigure[$T=0.60\,T_c$]{\includegraphics[width=0.42\textwidth]{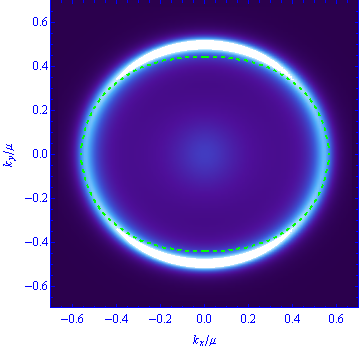}} \hfill
\subfigure[$T=0.43\,T_c$]{\includegraphics[width=0.42\textwidth]{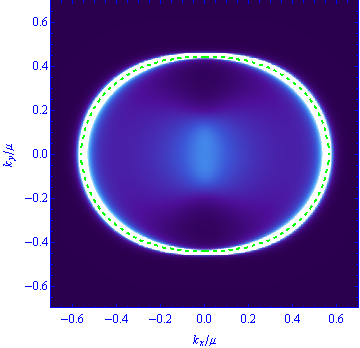}}\hfill
\subfigure[$T=0.20\,T_c$]{\includegraphics[width=0.42\textwidth]{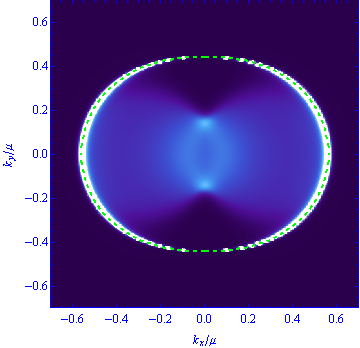}}\hfill
\subfigure[$T=0$]{\includegraphics[width=0.42\textwidth]{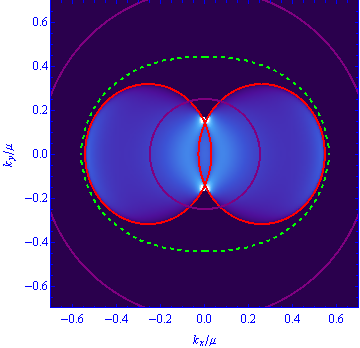}}
\caption{
The spectral function for different temperatures and fixed $\lambda=0.40<\lambda_c$, ${\omega} =0.25\,\mu$.
It is observed the presence of spherical symmetry at $T = T_c$.
The dashed lines indicates the locus of the normal modes. The purple and red circles represent the UV and IR Dirac cones respectively.
\label{fig:funcionespectrallambda04omega0251}
}
\end{figure}

\begin{figure}[H]
\centering
\subfigure[$T=T_{c}$]{\includegraphics[width=0.42\textwidth]{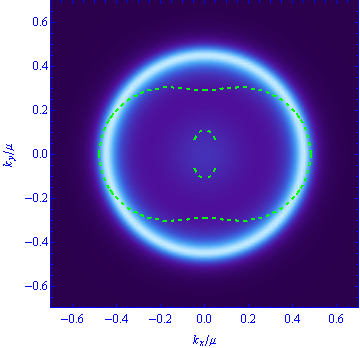}} \hfill
\subfigure[$T=0.85\,T_c$]{\includegraphics[width=0.42\textwidth]{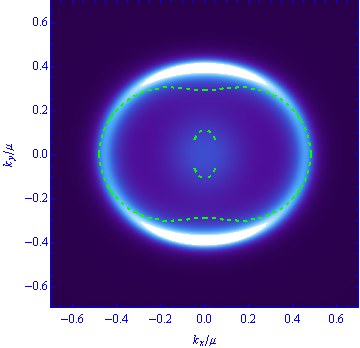}} \vfill
\subfigure[$T=0.60\,T_c$]{\includegraphics[width=0.42\textwidth]{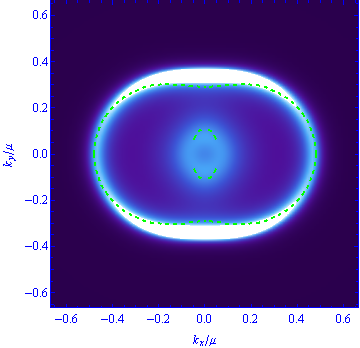}} \hfill
\subfigure[$T=0.43\,T_c$]{\includegraphics[width=0.42\textwidth]{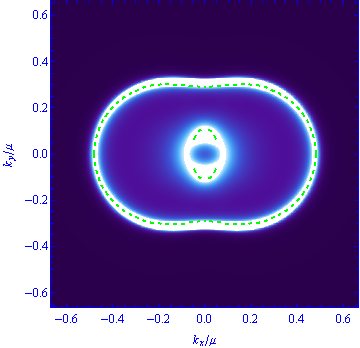}}\hfill
\subfigure[$T=0.20\,T_c$]{\includegraphics[width=0.42\textwidth]{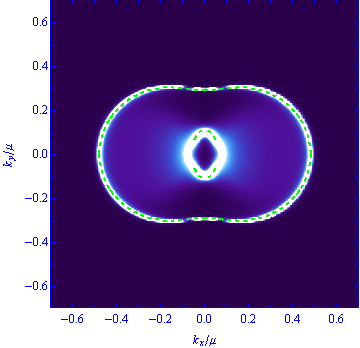}}\hfill
\subfigure[$T=0$]{\includegraphics[width=0.42\textwidth]{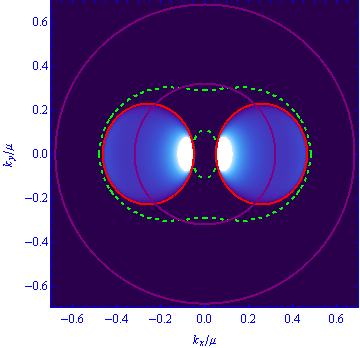}}
\caption{
The spectral function for different temperatures and fixed $\lambda=0.40<\lambda_c$, ${\omega}=0.18\,\mu$.
It is observed the presence of spherical symmetry at $T = T_c$.
The dashed lines indicates the locus of the normal modes. The purple and red circles represent the UV and IR Dirac cones respectively.
\label{fig:funcionespectrallambda04omega0181}
}
\end{figure}

\begin{figure}[H]
\centering
\subfigure[$T=T_{c}$]{\includegraphics[width=0.42\textwidth]{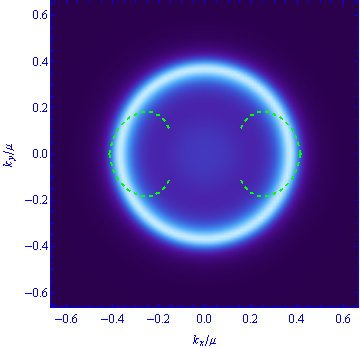}} \hfill
\subfigure[$T=0.85\,T_c$]{\includegraphics[width=0.42\textwidth]{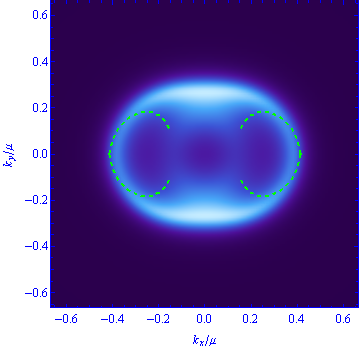}} \vfill
\subfigure[$T=0.60\,T_c$]{\includegraphics[width=0.42\textwidth]{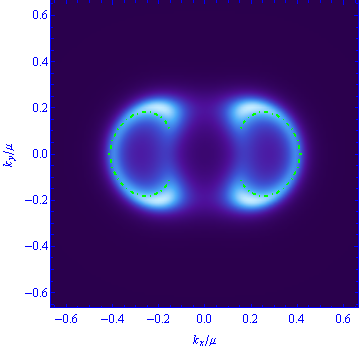}} \hfill
\subfigure[$T=0.43\,T_c$]{\includegraphics[width=0.42\textwidth]{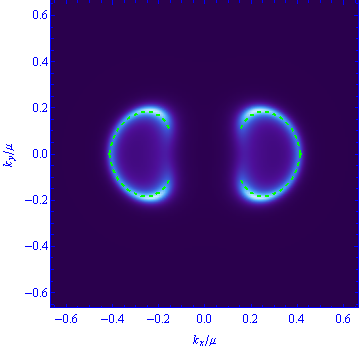}}\hfill
\subfigure[$T=0.20\,T_c$]{\includegraphics[width=0.42\textwidth]{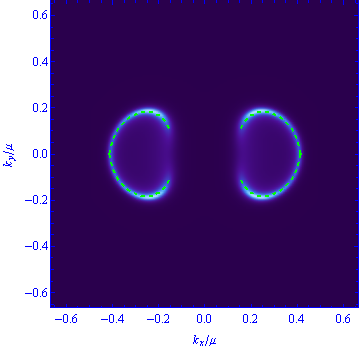}}\hfill
\subfigure[$T=0$]{\includegraphics[width=0.42\textwidth]{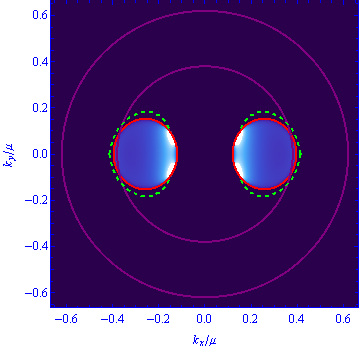}}
\caption{
The spectral function for different temperatures and fixed $\lambda=0.40<\lambda_c$, ${\omega}=0.12\,\mu$.
It is observed the presence of spherical symmetry at $T = T_c$.
The dashed lines indicates the locus of the normal modes. The purple and red circles represent the UV and IR Dirac cones respectively.
\label{fig:funcionespectrallambda04omega0121}}
\end{figure}

\begin{figure}[H]
\centering
\subfigure[$T=T_{c}$]{\includegraphics[width=0.42\textwidth]{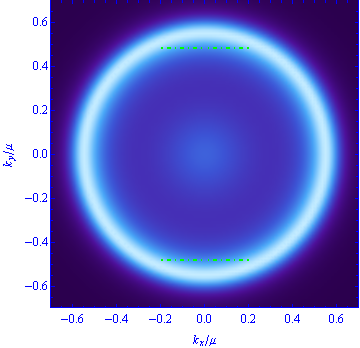}} \hfill
\subfigure[$T=0.79\,T_c$]{\includegraphics[width=0.42\textwidth]{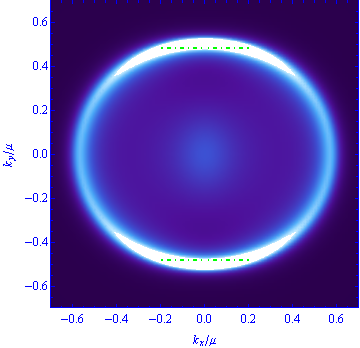}} \vfill
\subfigure[$T=0.60\,T_c$]{\includegraphics[width=0.42\textwidth]{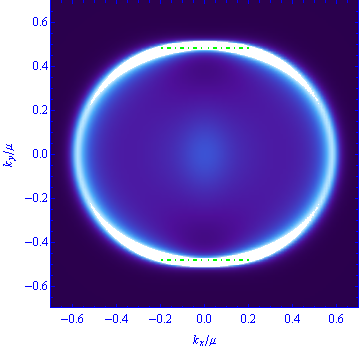}} \hfill
\subfigure[$T=0.38\,T_c$]{\includegraphics[width=0.42\textwidth]{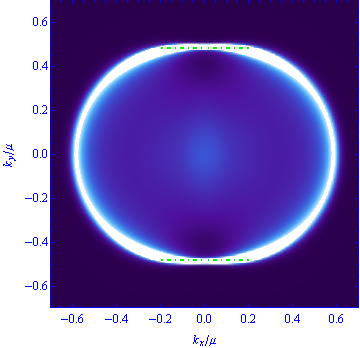}}\hfill
\subfigure[$T=0.28\,T_c$]{\includegraphics[width=0.42\textwidth]{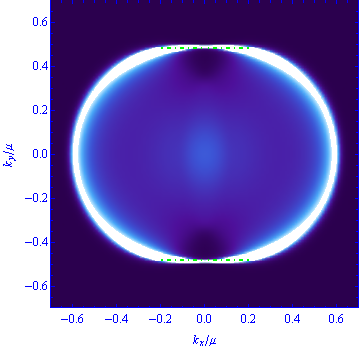}}\hfill
\subfigure[$T=0$]{\includegraphics[width=0.42\textwidth]{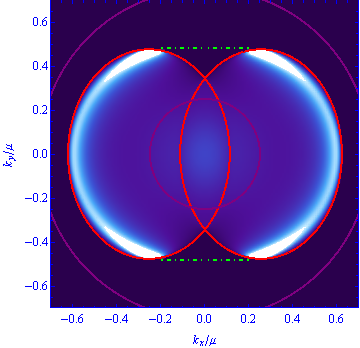}}
\caption{
The spectral function for different temperatures and fixed $\lambda=0.62=\lambda_c$, ${\omega} =0.25\mu$.
It is observed the presence of spherical symmetry at $T = T_c$.
The dashed lines indicates the locus of the normal modes. The purple and red circles represent the UV and IR Dirac cones respectively.
\label{fig:funcionespectrallambda062omega0251}}
\end{figure}

\begin{figure}[H]
\centering
\subfigure[$T=T_{c}$]{\includegraphics[width=0.42\textwidth]{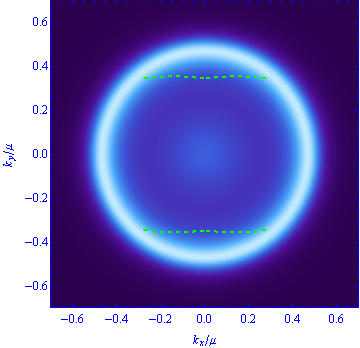}} \hfill
\subfigure[$T=0.79\,T_c$]{\includegraphics[width=0.42\textwidth]{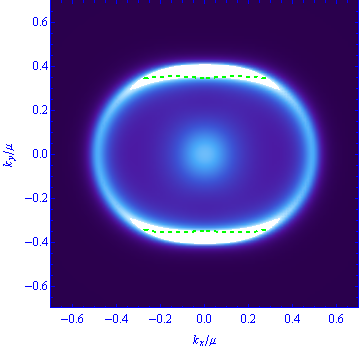}} \vfill
\subfigure[$T=0.60\,T_c$]{\includegraphics[width=0.42\textwidth]{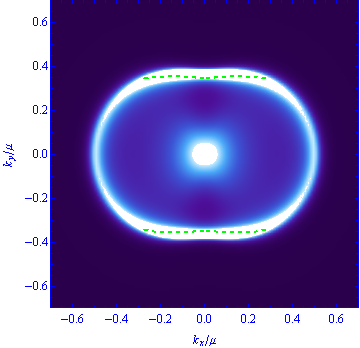}} \hfill
\subfigure[$T=0.38\,T_c$]{\includegraphics[width=0.42\textwidth]{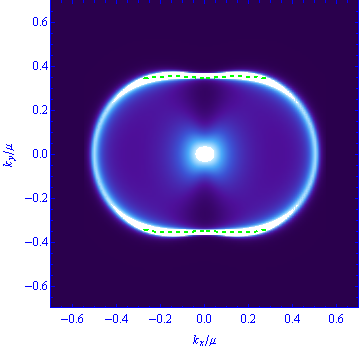}}\hfill
\subfigure[$T=0.28\,T_c$]{\includegraphics[width=0.42\textwidth]{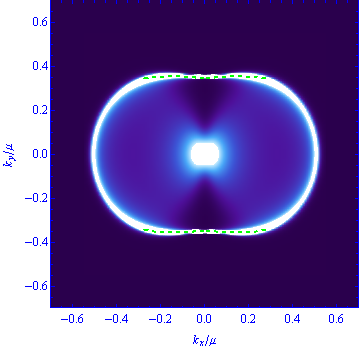}}\hfill
\subfigure[$T=0$]{\includegraphics[width=0.42\textwidth]{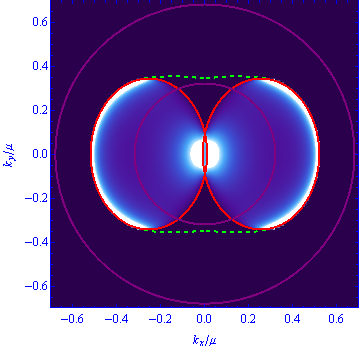}}
\caption{
The spectral function for different temperatures and fixed $\lambda=0.62=\lambda_c$, ${\omega} =0.18\,\mu$.
It is observed the presence of spherical symmetry at $T = T_c$.
The dashed lines indicates the locus of the normal modes. The purple and red circles represent the UV and IR Dirac cones respectively.
\label{fig:funcionespectrallambda062omega0181}}
\end{figure}

\begin{figure}[H]
\centering
\subfigure[$T=T_{c}$]{\includegraphics[width=0.42\textwidth]{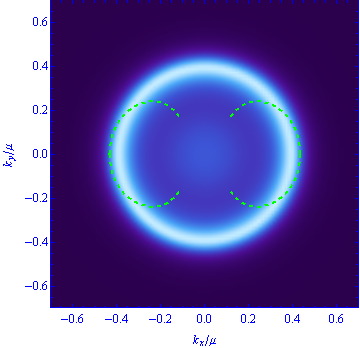}} \hfill
\subfigure[$T=0.79\,T_c$]{\includegraphics[width=0.42\textwidth]{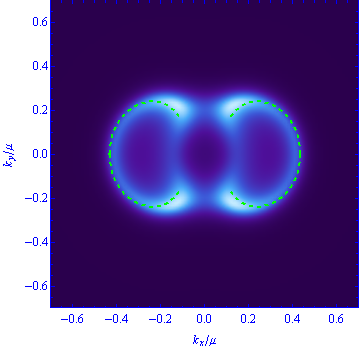}} \vfill
\subfigure[$T=0.60\,T_c$]{\includegraphics[width=0.42\textwidth]{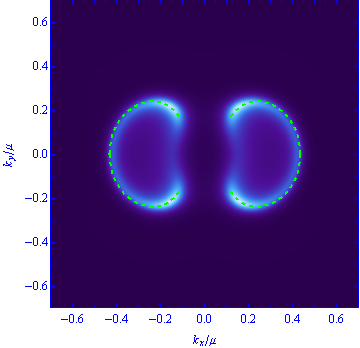}} \hfill
\subfigure[$T=0.38\,T_c$]{\includegraphics[width=0.42\textwidth]{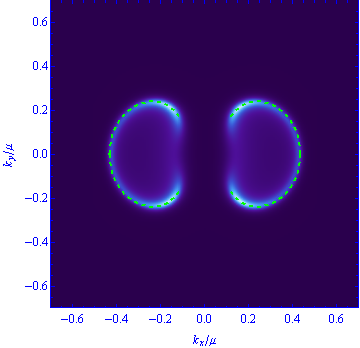}}\hfill
\subfigure[$T=0.28\,T_c$]{\includegraphics[width=0.42\textwidth]{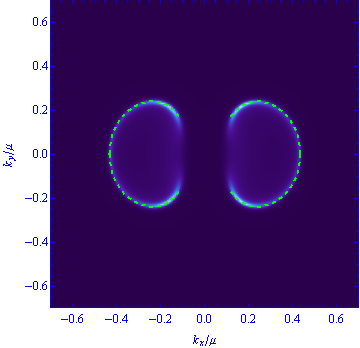}}\hfill
\subfigure[$T=0$]{\includegraphics[width=0.42\textwidth]{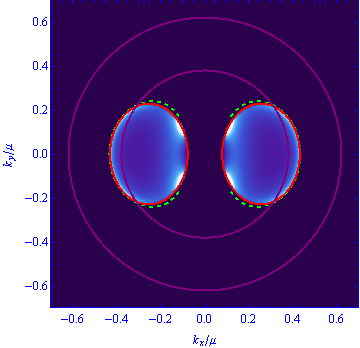}}
\caption{
The spectral function for different temperatures and fixed $\lambda=0.62=\lambda_c$, ${\omega} =0.12\,\mu$.
It is observed the presence of spherical symmetry at $T = T_c$.
The dashed lines indicates the locus of the normal modes. The purple and red circles represent the UV and IR Dirac cones respectively.
\label{fig:funcionespectrallambda062omega0121}
}
\end{figure}

\begin{figure}[H]
\centering
\subfigure[$T=0.80\,T_c$]{\includegraphics[width=0.32\textwidth]{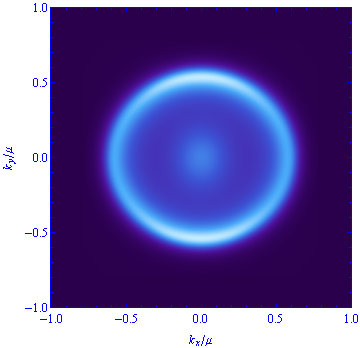}}
\subfigure[$T=0.40\,T_c$]{\includegraphics[width=0.32\textwidth]{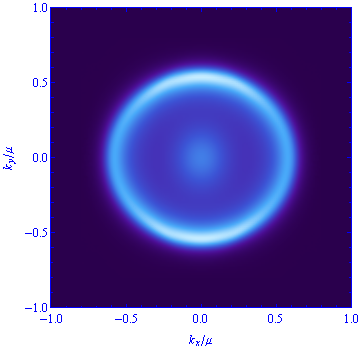}}
\subfigure[$T=0$]{\includegraphics[width=0.32\textwidth]{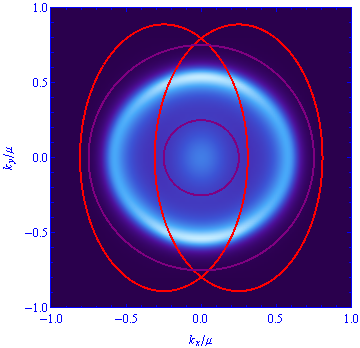}}
\vfill
\caption{
The spectral function for different temperatures and fixed $\lambda=0.80>\lambda_c$, ${\omega} =0.25\,\mu$.
It remains almost unchanged for the entire range of the temperatures.
The purple and red circles represent the UV and IR Dirac cones respectively.
\label{fig:funcionespectrallambda08omega0251}}
\end{figure}

\begin{figure}[H]
\centering
\subfigure[$T=0.80\,T_c$]{\includegraphics[width=0.32\textwidth]{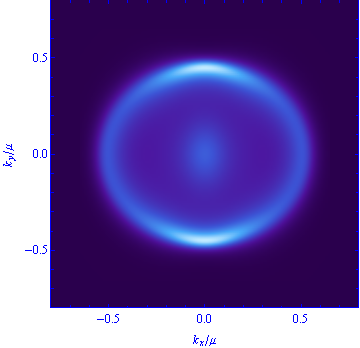}}
\subfigure[$T=0.40\,T_c$]{\includegraphics[width=0.32\textwidth]{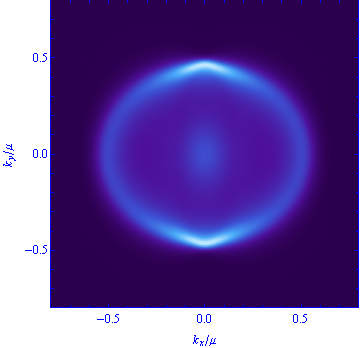}}
\subfigure[$T=0$]{\includegraphics[width=0.32\textwidth]{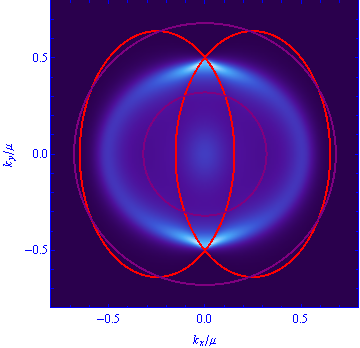}}
\vfill
\caption{
The spectral function for different temperatures and fixed $\lambda=0.80>\lambda_c$, ${\omega}=0.18\,\mu$.
It remains almost unchanged for the entire range of the temperatures.
The purple and red circles represent the UV and IR Dirac cones respectively.
\label{fig:funcionespectrallambda08omega0181}}
\end{figure}

\begin{figure}[H]
\centering
\subfigure[$T=0.80\,T_c$]{\includegraphics[width=0.32\textwidth]{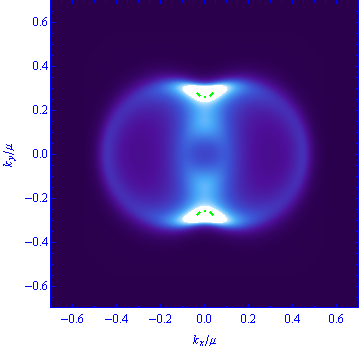}}
\subfigure[$T=0.40\,T_c$]{\includegraphics[width=0.32\textwidth]{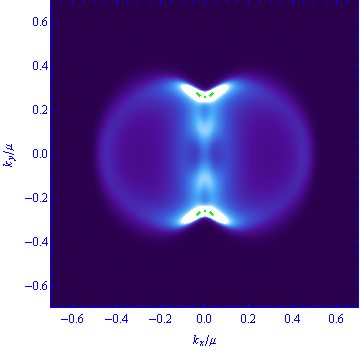}}
\subfigure[$T=0$]{\includegraphics[width=0.32\textwidth]{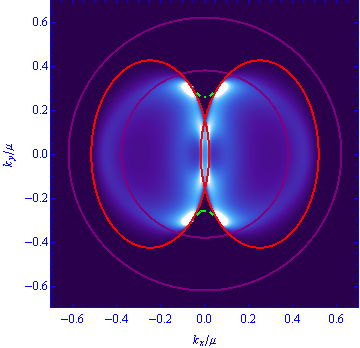}}
\vfill
\caption{
The spectral function for different temperatures and fixed $\lambda=0.80>\lambda_c$, ${\omega} =0.12\,\mu$.
It remains almost unchanged for the entire range of the temperatures.
The purple and red circles represent the UV and IR Dirac cones respectively.
\label{fig:funcionespectrallambda08omega0121}}
\end{figure}

Figures \ref{fig:funcionespectrallambda04omega0251} to \ref{fig:funcionespectrallambda08omega0121} show the evolution of the spectral function as a function of the temperature. Each figure corresponds to a fixed $\lambda$ and $\omega$. In all cases, as the temperature decreases starting from the critical temperature, the spherical symmetry is rapidly lost, the local maxima become sharper and approach the locus of the normal modes. They lie, as showed in reference \cite{Gubser:2010dm}, between the UV and IR Dirac cones.
Furthermore, the zero temperature plot shows the support of the spectral function completely contained inside the IR Dirac cones, in agreement with discussions in subsection \ref{sec:fermions-in-the-QFT} and appendix A.

The sets formed for figures \ref{fig:funcionespectrallambda04omega0251} to  \ref{fig:funcionespectrallambda04omega0121},
figures \ref{fig:funcionespectrallambda062omega0251} to \ref{fig:funcionespectrallambda062omega0121},
and figures \ref{fig:funcionespectrallambda08omega0251} to \ref{fig:funcionespectrallambda08omega0121} respectively
correspond to three fixed values of $\lambda$, smaller, equal and bigger than $\lambda_c$.
Each figure in the set shows plots of the spectral function for different temperatures at a fixed value of $\omega$.
The value of $\omega$ decreases from figure to figure inside each set.
It is evident by comparing figures corresponding to the same temperature inside a single set,
that the support of the spectral function in momentum space gets smaller as $\omega$ decreases for fixed $\lambda$.
By comparing different sets we see that the dependence on temperature and frequency of the spectral function is milder for
$\lambda>\lambda_c$ than in the cases $\lambda \leq \lambda_c$. Notice also that even if spherical symmetry is absent, it seems to be recovered as the frequency is increased. For $\lambda>\lambda_c$ and high frequencies, figure \ref{fig:funcionespectrallambda08omega0251}, normal modes does not exist. This is in agreement with the absence of a region between the IR and the UV Dirac cones. Instead, in figure \ref{fig:funcionespectrallambda08omega0121}, two small, open and disconnected regions are present, where normal modes show up.

\begin{figure}[H]
\begin{center}
\begin{minipage}[t]{\textwidth}
\begin{minipage}{0.5\textwidth}
\center
\subfigure[$\omega=0$, $T=T_c$]
{\includegraphics[width=0.65 \textwidth]{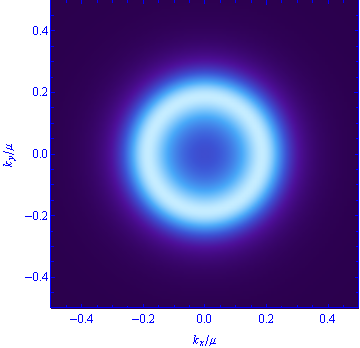}}
\subfigure[$\omega=0$, $T=0.95\,T_c$]
{\includegraphics[width=0.65 \textwidth]{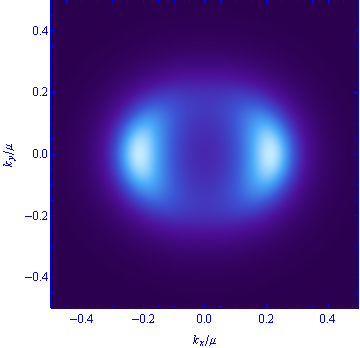}}
\subfigure[$\omega=0$, $T=0.85\,T_c$]{\includegraphics[width=0.65 \textwidth]{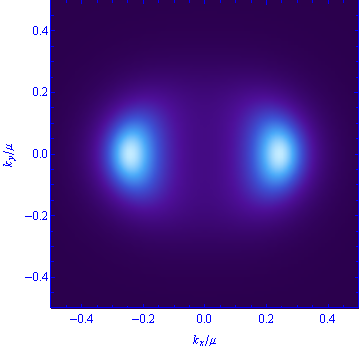}}
\subfigure[$\omega=0$, $T=0.20\,T_c$]{\includegraphics[width=0.65 \textwidth]{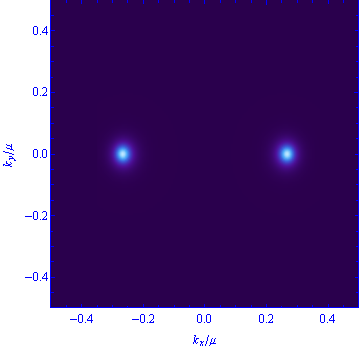}}
\end{minipage}
\renewcommand{\thesubfigure}{(\arabic{subfigure})}
\makeatletter
\makeatother
\setcounter{subfigure}{0}
\hfill
\begin{minipage}{0.44\textwidth}
\center \vspace{-0.5cm}
\subfigure[][]{\includegraphics[width=0.96\textwidth]{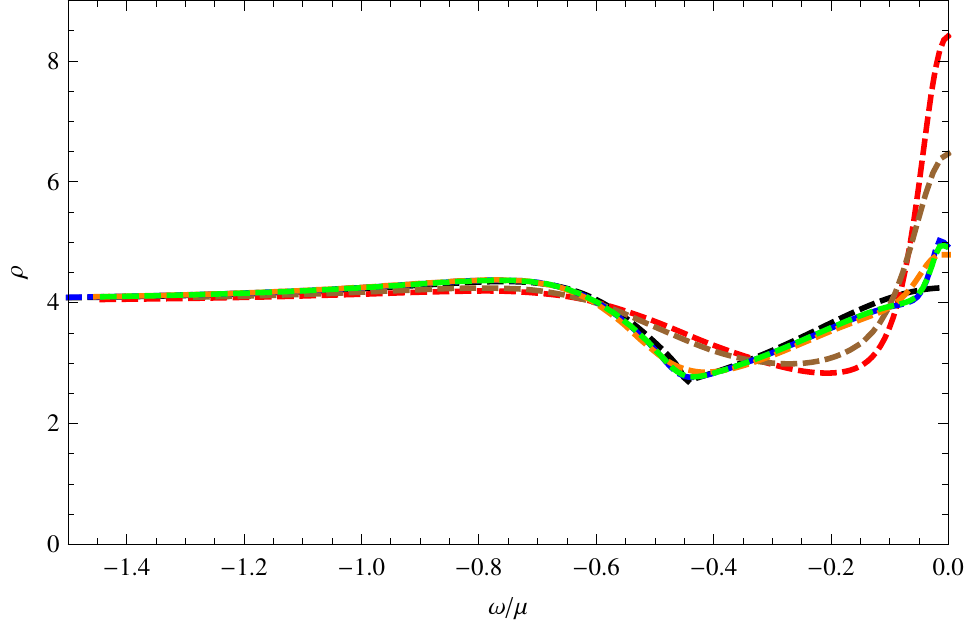}}
\subfigure[][]{\includegraphics[width=0.98\textwidth]{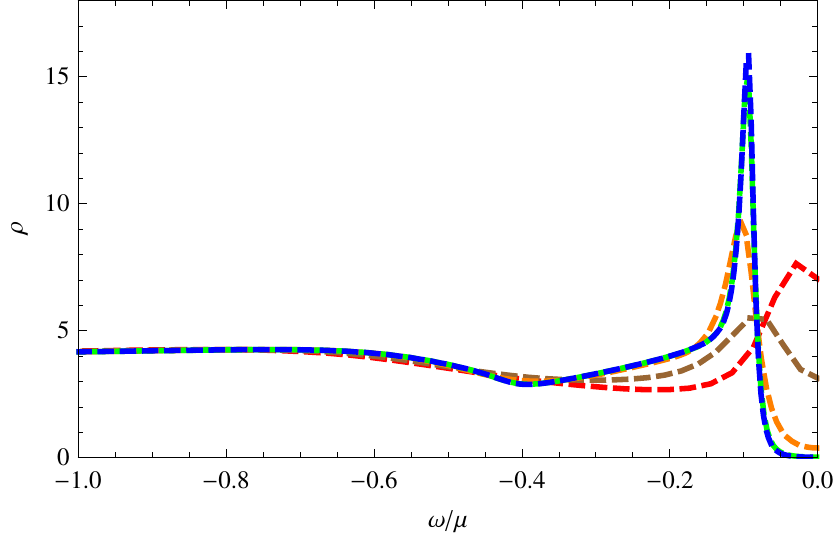}}
\subfigure[][]{\includegraphics[width=0.98\textwidth]{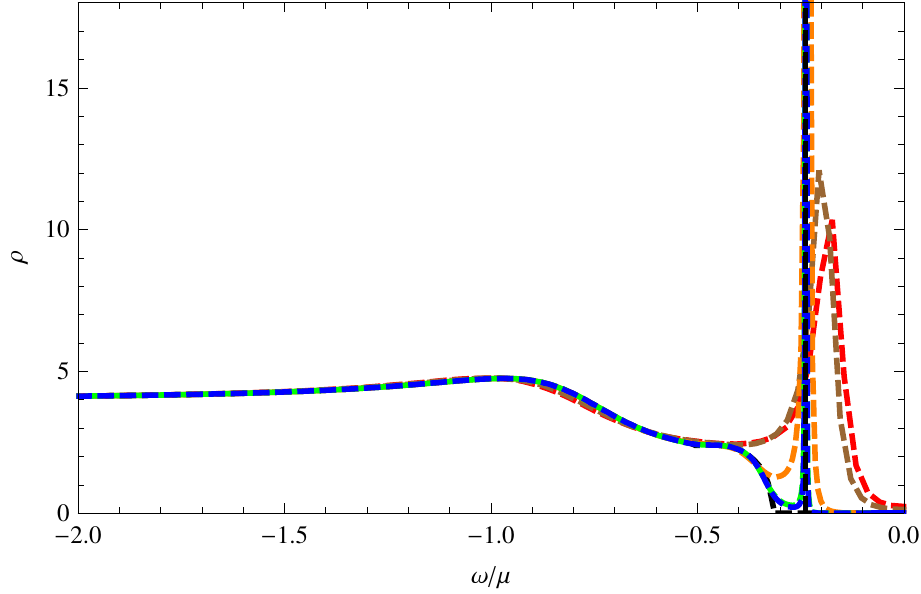}}
\end{minipage}
\end{minipage}
\caption{
The spectral function at fixed $\lambda = 0.40<\lambda_c$.
Left: The spectral surface $\omega = 0$ at different temperatures.
Right:
(1) $\rho$ as a function of $\omega$ at $(k_x,k_y)=(k_*, 0)$;
(2) $\rho$ as a function of $\omega$ at $(k_x,k_y)=(0.20,0.10)\mu$;
(3) $\rho$ as a function of $\omega$ at $(k_x,k_y)=(0.20,0.40)\mu$.
Each graph in these plots corresponds to the temperatures: $T=T_{c}\,$ (red), $T=0.97\,T_{c}\,$ (brown), $T=0.86\,T_{c}\,$ (orange), $T=0.43\,T_{c}\,$ (green), $T=0.21\,T_{c}\,$ (blue),  $T=0$ (black).
\label{evolucionlambda04variosk}
}
\end{center}
\end{figure}

\begin{figure}[H]
\begin{center}
\begin{minipage}[t]{\textwidth}
\begin{minipage}{0.5\textwidth}
\center
\subfigure[$\omega=0$, $T=T_c$]
{\includegraphics[width=0.65 \textwidth]{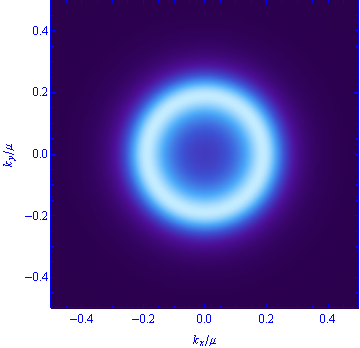}}
\subfigure[$\omega=0$, $T=0.97\,T_c$]
{\includegraphics[width=0.65 \textwidth]{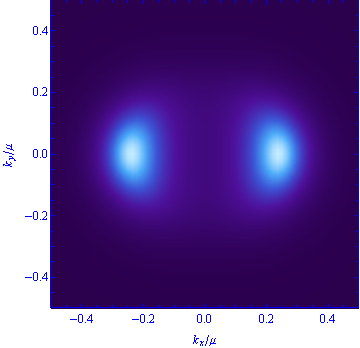}}
\vfill
\subfigure[$\omega=0$, $T=0.79\,T_c$]
{\includegraphics[width=0.65 \textwidth]{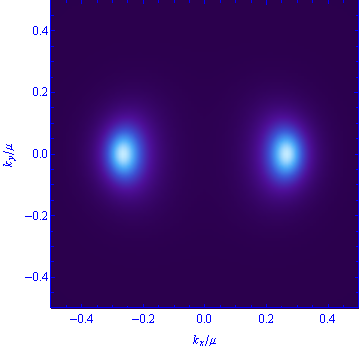}}
\subfigure[$\omega=0$, $T=0.41\,T_c$]
{\includegraphics[width=0.65 \textwidth]{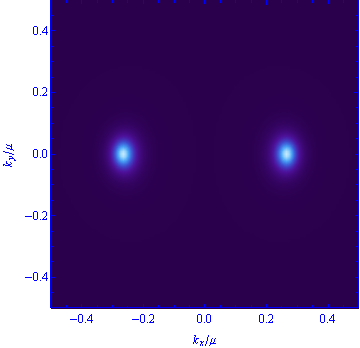}}
\end{minipage}
\renewcommand{\thesubfigure}{(\arabic{subfigure})}
\makeatletter
\makeatother
\setcounter{subfigure}{0}
\hfill
\begin{minipage}{0.44\textwidth}
\center \vspace{-0.5cm}
\subfigure[][]{\includegraphics[width=0.98\textwidth]{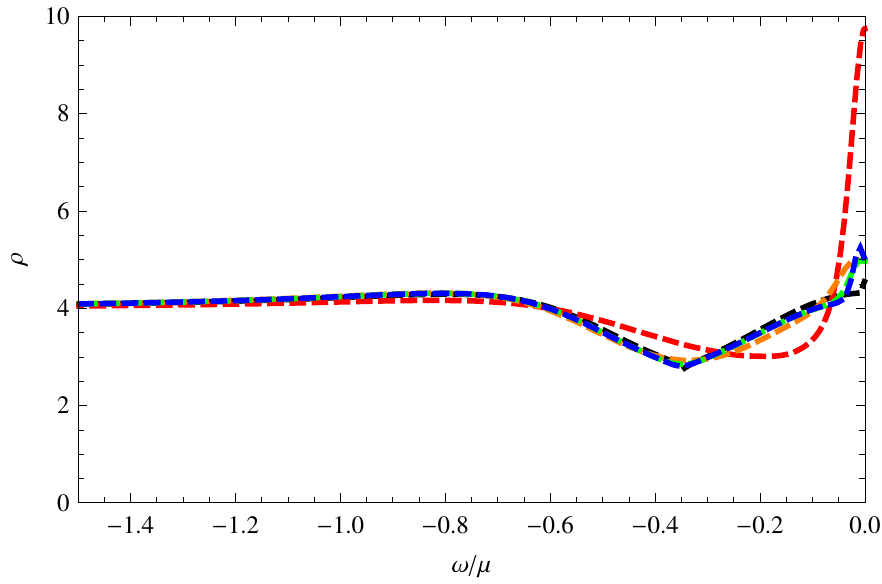}}
\subfigure[][]{\includegraphics[width=0.98\textwidth]{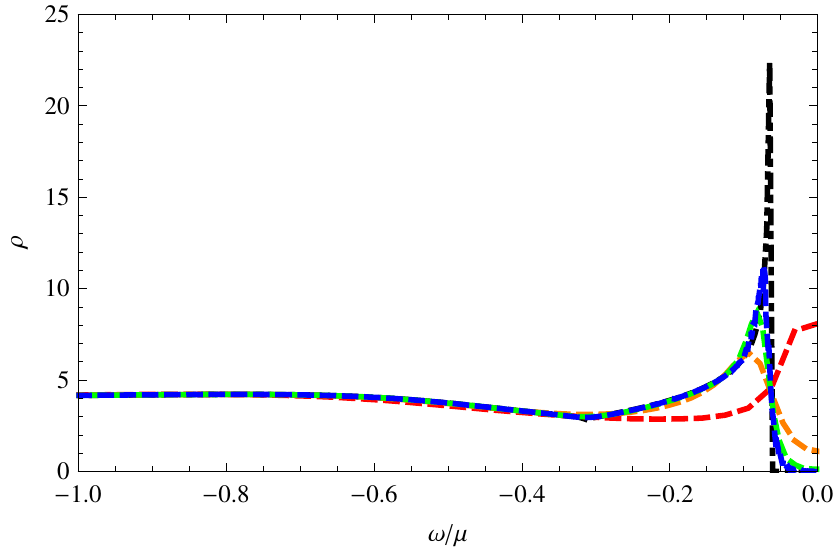}}
\subfigure[][]{\includegraphics[width=0.98\textwidth]{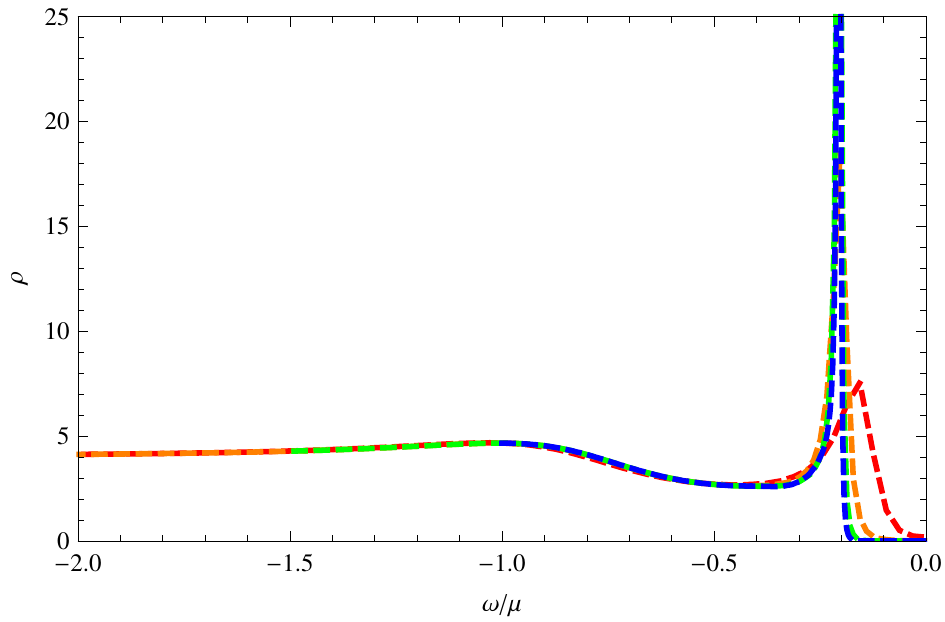}}
\end{minipage}
\end{minipage}
\caption{
The spectral function at fixed $\lambda=\lambda_c = 0.62$.
Left:
The spectral surface $\omega=0$ at different temperatures.
Right:
(1) $\rho$ as a function of $\omega$ at $(k_x,k_y)= (k_*, 0)$;
(2) $\rho$ as a function of $\omega$ at $(k_x,k_y)=(0.20,0.10)\mu$;
(3) $\rho$ as a function of $\omega$  at $(k_x,k_y) =(0.20,0.40)\mu$.
Each graph in these plots corresponds to the temperatures:
$T=T_c\,$ (red), $T=0.97\,T_c\,$ (purple) $T=0.76\,T_c\,$ (orange), $T=0.38\,T_c\,$ (green),
$T=0.28\,T_c\,$ (blue), $T=0\,$ (black).
\label{evolucionlambda062variosk}}
\end{center}
\end{figure}

\begin{figure}[H]
\begin{center}
\begin{minipage}[t]{\textwidth}
\begin{minipage}{0.5\textwidth}
\center
\subfigure[$\omega=0$, $T=T_c$]
{\includegraphics[width=0.65 \textwidth]{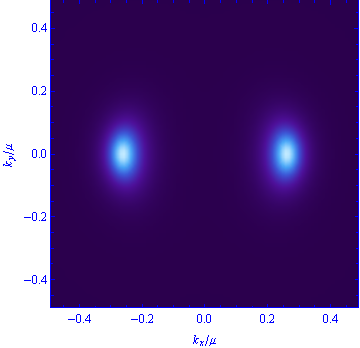}}
\subfigure[$\omega=0$, $T=0.80\,T_c$]
{\includegraphics[width=0.65 \textwidth]{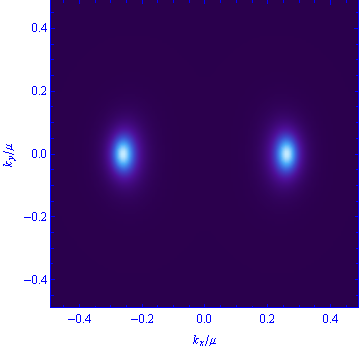}}
\vfill
\vfill
\subfigure[$\omega=0$, $T=0.40\,T_c$]
{\includegraphics[width=0.65 \textwidth]{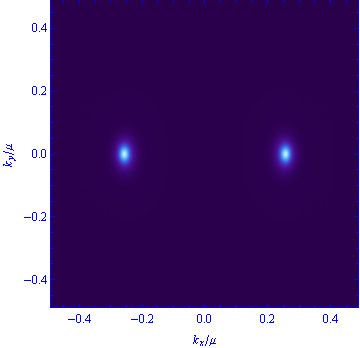}}
\end{minipage}
\renewcommand{\thesubfigure}{(\arabic{subfigure})}
\makeatletter
\makeatother
\setcounter{subfigure}{0}
\hfill
\begin{minipage}{0.44\textwidth}
\center \vspace{-0.5cm}
\subfigure[][]
{\includegraphics[width=0.98\textwidth]{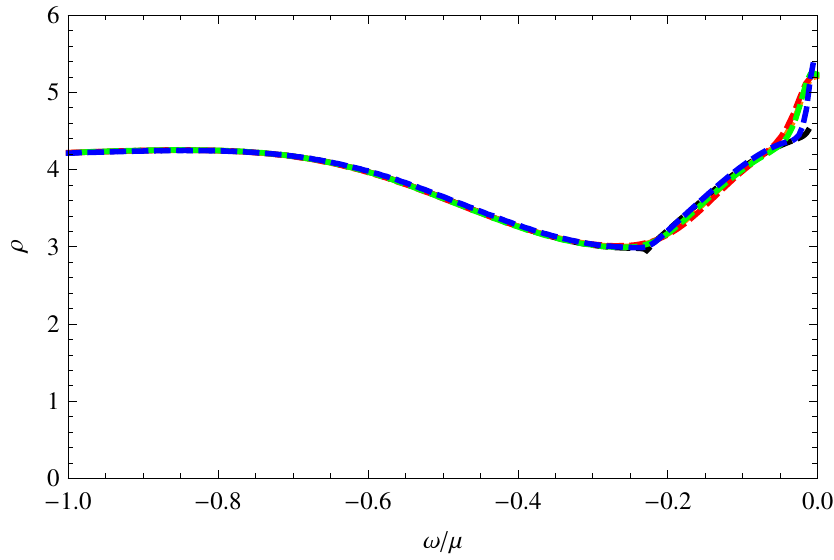}}
\subfigure[][]
{\includegraphics[width=0.98\textwidth]{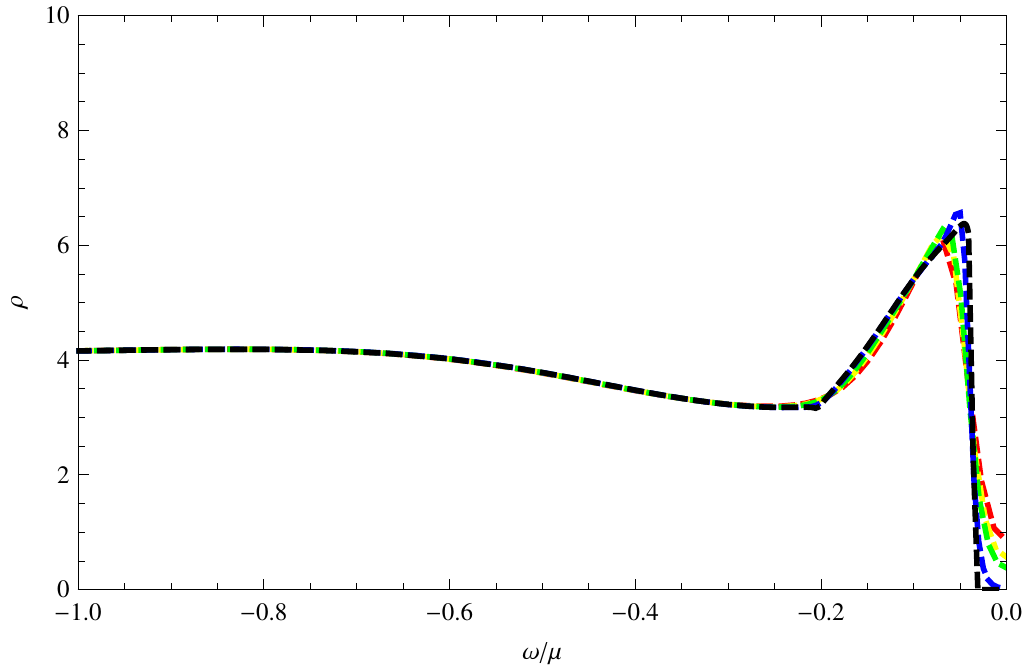}}
\subfigure[][]
{\includegraphics[width=0.96\textwidth]{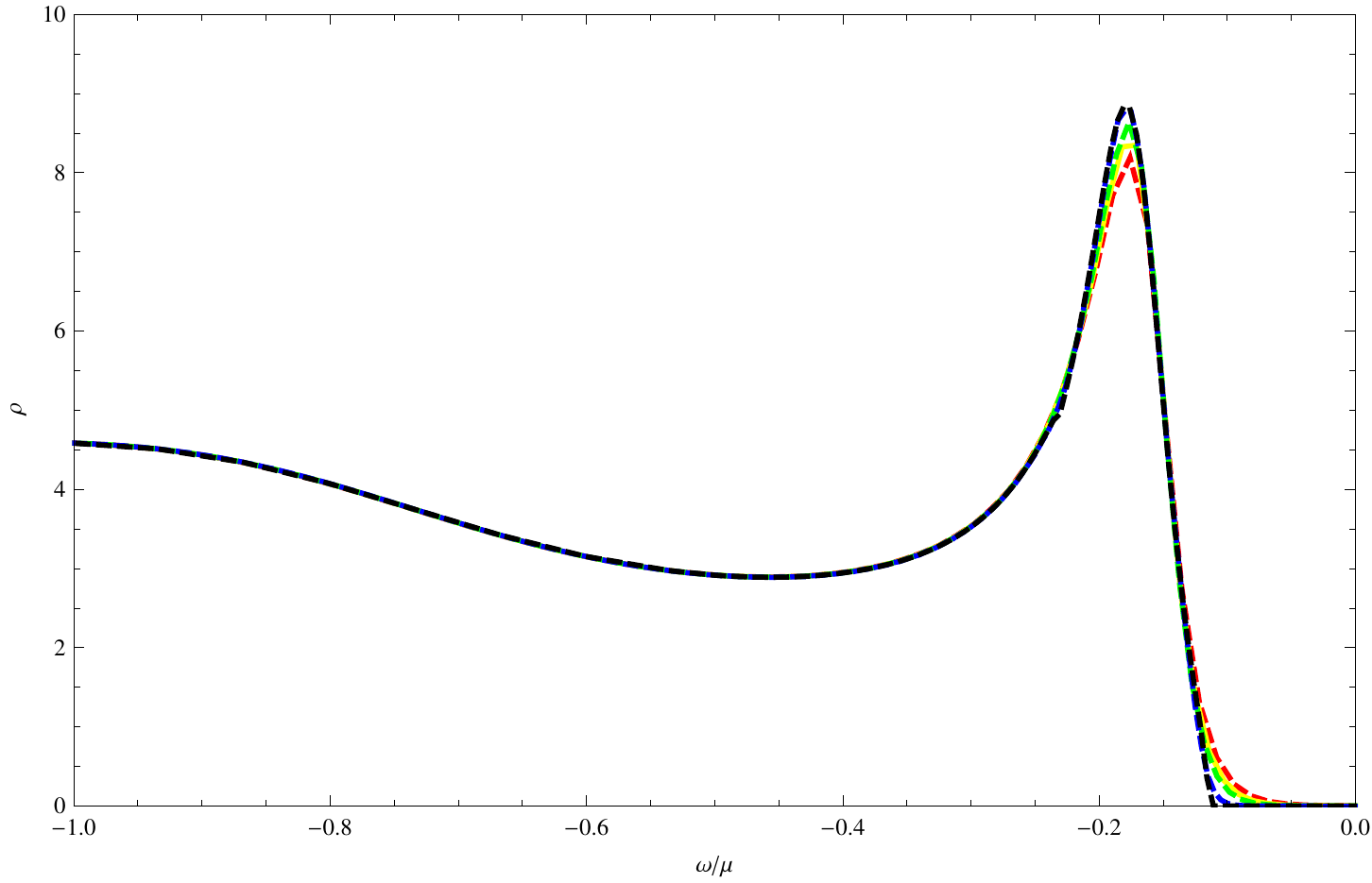}}
\end{minipage}
\end{minipage}
\caption{
The spectral function at fixed $\lambda = 0.80>\lambda_c$.
Left: The spectral surface $\omega = 0$ at different temperatures.
Right:
(1) $\rho$ as a function of $\omega$  at $(k_x,k_y)/=(k_*, 0)$;
(2) $\rho$ as a function of $\omega$  at ${(k_x,k_y)}/{\mu}=(0.20,0.10)$;
(3) $\rho$ as a function of $\omega$  at ${(k_x,k_y)}/{\mu}=(0.20,0.40)$.
Each graph in these plots corresponds to the temperatures: $T=T_{c}\,$ (red), $T=0.90\,T_{c}\,$ (orange),
$T=0.90\, T_{c}$ (yellow), $T=0.80\,T_{c}\,$ (green), $T=0.40\,T_{c}\,$ (blue), $T=0\,$  (black).
\label{evolucionlambda08variosk}
}
\end{center}
\end{figure}

Figures \ref{evolucionlambda04variosk} to  \ref{evolucionlambda08variosk} correspond to three different values of $\lambda$,
smaller, equal and bigger than $\lambda_c$.
On the left, the spectral surface $\omega=0$ is shown for different temperatures. On the right, the spectral function as a function of the frequency is plotted for several temperatures, for three different momenta. Of these, subfigures (1) are plotted  at the momentum $(k_x,k_y)_* = (k_*, 0)$ that corresponds to one of the two maxima of the spectral surface at each temperature. Defining the Fermi momentum $k_F$ as the location of one of the maxima when the temperature vanishes $T=0$, the Fermi surface consists of the two points $\pm(k_x^F,k_y^F) = (\pm k_F, 0)$. It is observed that limit of the spectral function as $\omega\to 0$ is bigger as the temperature approaches $T_c$. It is also observed that the peaks in the spectral function as a function of $\omega$ in subfigures (2) (3)
move to the left and get sharpen as the temperature decreases, corresponding at $T=0$ to the appearance of the normal modes.

%

%

%

%

%

%

%

%

%

%

%

%

\newpage

\section{Discussion}
\label{sec:discussion}
In this paper we have studied spectral functions of fermionic systems at zero temperature/vanishing entropy and at finite temperature and entropy in back-reacted $p$-wave superconductors through the computation of fermionic correlation functions via the AdS/CFT correspondence, extending previous work in the literature.

The results show a breaking of rotational invariance that gets stronger as the temperature is decreased.
The temperature dependence is nevertheless milder as the gravitational coupling is increased.
The support of the spectral function gets reduced as the temperature is decreased, the peaks at the normal modes becoming sharper.
At zero temperature, the support is completely contained inside the IR Dirac cones and in the sharp divergences at the normal modes,
the last ones representing the Fermi surface when $\omega$ goes to zero.
Since the normal modes exist inside the UV Dirac cones and outside the IR ones, this behavior is consistent with the appearance of Fermi arcs whenever the intersection of these regions does not allow for a closed Fermi surface.

As previously reported (c.f. \cite{Faulkner:2010da}) the behavior of the spectral function at zero temperature as a function of the frequency is not sharp enough to allow for an interpretation in terms of quasiparticles, as is expected for a strongly interacting system.
In spite of that, it is interesting to notice that some of the numerical results obtained can be intuitively understood in terms of level filling, as follows.
At zero temperature, normal modes set a relation between frequency and momenta that mimics a quasiparticle dispersion relation.
The level surfaces of such dispersion relation for different values of $\omega/\mu$ at different $\lambda$ are plotted
(with a dotted line) in figures \ref{fig:funcionespectrallambda04omega0251}-\ref{fig:funcionespectrallambda062omega0121} (f)
and \ref{fig:funcionespectrallambda08omega0251}-\ref{fig:funcionespectrallambda08omega0121} (c) as the location of the normal modes.
For high values of $\omega/\mu$ (figures \ref{fig:funcionespectrallambda04omega0251} and \ref{fig:funcionespectrallambda062omega0251})
such level surfaces are rotationally invariant and approach the UV Dirac cones.
On the other hand, for low values of $\omega/\mu$ (figures \ref{fig:funcionespectrallambda04omega0121} and \ref{fig:funcionespectrallambda062omega0121}) rotational invariance is broken and the level surfaces approach the IR Dirac cones.
As $\omega/\mu$ decreases, these lines can be understood as the Fermi surfaces of the system, corresponding to increasing values of $\mu$.
As the quotient $T/\mu$ is increased starting from zero, this can be understood as rising the temperature or lowering the chemical potential.
Now, in a nearly free fermionic system, a rising temperature naturally gives origin to a thickening of the Fermi surface from Fermi-Dirac statistics, while lowering the chemical potential fills the states in momentum space up to  a level curve of the dispersion relation corresponding to a lower frequency.
Now, we can see in the plots that this is exactly what happens: as we rise the temperature, the spectral function divergence at the normal modes gets milder, wider in momentum space, and it gets deformed into the form of the normal modes corresponding to a higher frequency in $\mu$ units, that is a lower $\mu$, see for example figures \ref{fig:funcionespectrallambda04omega0181} (b) and \ref{fig:funcionespectrallambda04omega0251} (f).
This fact can be presumably explained by following the analysis of single-trace operators correlation functions and the behavior of them as free fields in the large $N$-limit that is certainly present in the holographic set-up, see reference \cite{Arsiwalla:2010bt} for details.

\newpage
\appendix
\section{About the positivity of the spectral function}
\label{sec:appendix-postivity}

In this appendix we establish the non-negative property of the spectral function following
the analysis of the $T=0$ case in \cite{Gubser:2010dm} and extending the result for the $T>0$ case.

The starting point is the (on-shell) conserved current
\be
J^A(x,z) \equiv i\,\bar\Psi(x;z)\;\Gamma^A\;\Psi(x;z)\quad\longrightarrow\quad
J^A(z;k) = i\frac{z^3}{s\,\sqrt{f}}\,\bar\psi_k(z)\;\Gamma^A\;\psi_k(z)
\ee
where on the right we used our ansatz introduced after (\ref{eq:eom-fermionic}).
The conservation equation reads,
\bea\label{eq:Q}
D_A J^A(z;k)&=&\frac{z^4}{s}\;Q'(z;k)=0\cr
Q(z;k)&\equiv& -2\,Im\left({\bar\psi^+}_k(z)\;\psi^-_k(z)\right) =
-2\,Im\left(\vec u_k(z){}^\dagger\,{\bf C}\,\vec v_k(z)\right) =
-2\,Im\left(\vec u_k(z){}^\dagger\,{\bf S}(z)^{-1}\,\vec u'_k(z)\right)
\eea
Here the matrix ${\bf S}(z)\equiv -i\,{\bf U}(z)\,{\bf C}$ is hermitian, $\,{\bf S}(z)^\dagger={\bf S}(z)$, and in the second equality we have introduced the notation (\ref{eq:four-tuples}) with the matrix $\bf C$ giving in (\ref{eq:action-fermionic}), while that in the last equality we used the equations of motion (\ref{eq:eom-four-tuples}).

From (\ref{eq:Q}) we conclude that, when evaluated on-shell, $Q(z;k)=Q(k)\,$ is a conserved quantity, and even if we do not know the explicit solution to the equations of motion we can evaluate it in known limits.

In first term we consider the UV limit; with the help of the asymptotic behavior (\ref{eq:solution-evolutionUV}), the relation (\ref{eq:M}) and the definition (\ref{eq:GMC}) we obtain
\be
Q_{UV}(k)\equiv Q(0;k) = i\, \left( {\bf C}\,\vec u_k^{(UV)}\right)^\dagger\;
\left( {\bf G}_R(k) - {\bf G}_R(k){}^\dagger \right)\;{\bf C}\,\vec u_k^{(UV)}
\ee
If we assume that $Q_{UV}(k)\geq 0$, and being the vector $\vec u_k^{(UV)}$ arbitrary,
we must conclude that the hermitian matrix
$\;i\,\left( {\bf G}_R(k) - {\bf G}_R(k){}^\dagger \right)$ is non-negative.
From the definition (\ref{eq:rho}) we conclude that,
\be
\rho(\omega;k_x,k_y) \equiv Tr \frac{i}{2}\,\left( {\bf G}_R(k) - {\bf G}_R(k){}^\dagger \right)\geq 0
\ee
being strictly zero iff all the eigenvalues are zero and then $Q_{UV}=0$

In second term we consider the IR limit; in this case we have to distinguish two cases.
When the entropy vanishes the asymptotic behavior for $z\rightarrow\infty$ is of the form
\be
{\vec u}_k(z) \simeq
e^{-\kappa_+\,z}\;\left(\begin{array}{cccc}b_1\\b_2\\-b_1\\-b_2\end{array}\right)+ e^{-\kappa_-\,z}\;\left(\begin{array}{cccc}b_3\\b_4\\b_3\\b_4\end{array}\right)
\label{eq:comportamiento-IR_ibc S=0-app}
\ee
where the vector $\vec b^t=(b_1, b_2, b_3, b_4)$ is read from (\ref{eq:comportamiento-IR_ibc S=0}) in terms of $\vec u_k^{(ir)}$
but such relation is not relevant here.
From (\ref{eq:Q}) we get
\bea
Q_{IR}(k) &\equiv& \!\!Q(z;k)|_{z\rightarrow\infty}\cr
&& \!\!\!\!\!\!\!\!\!\!\!\!\!\!\!\!\!\!\!\!\!\!\!\!=\lim_{z\rightarrow\infty}
\left(
2\,Im(\kappa_+)\,\left| e^{-\kappa_+\,z} \right|^2\;
\left(\begin{array}{cccc}b_1\\b_2\\-b_1\\-b_2\end{array}\right)^\dagger\!{\bf S}_{IR}^{-1}\!
\left(\begin{array}{cccc}b_1\\b_2\\-b_1\\-b_2\end{array}\right) +
2\,Im(\kappa_-)\,\left| e^{-\kappa_-\,z} \right|^2\;
  \left(\begin{array}{cccc}b_3\\b_4\\b_3\\b_4\end{array}\right)^\dagger\!{\bf S}_{IR}^{-1}\!
\left(\begin{array}{cccc}b_3\\b_4\\b_3\\b_4\end{array}\right)\right.\cr
&&\!\!\!\!\!\!\!\!\!\!\!\!\!\!\!\!\!\!\!\!\!\!\!\!+\left.2\,Im\left(\kappa_+\, e^{-(\kappa_+ +\kappa_-{}^*)\,z}\;
\left(\begin{array}{cccc}b_3\\b_4\\b_3\\b_4\end{array}\right)^\dagger\;{\bf S}_{IR}^{-1}\;
\left(\begin{array}{cccc}b_1\\b_2\\-b_1\\-b_2\end{array}\right) +
\kappa_-\, e^{-(\kappa_+{}^* +\kappa_-)\,z}\;
\left(\begin{array}{cccc}b_1\\b_2\\-b_1\\-b_2\end{array}\right)^\dagger\;{\bf S}_{IR}^{-1}\;
\left(\begin{array}{cccc}b_3\\b_4\\b_3\\b_4\end{array}\right)
\right)\right)\cr\label{eq:QIR}
&&
\eea
where
\be
{\bf S}_{IR}\equiv {\bf S}|_{z\rightarrow\infty} =
\left(\begin{array}{cccc}{\bf S_1}&{\bf S_2}\\{\bf S_2}&{\bf S_1}\end{array}\right)\qquad;\qquad
{\bf S_1}\equiv -\frac{\omega}{s_0} {\bf 1} - \frac{k_y}{g_0}\,{\bf\sigma_1}+ g_0\,k_x\,{\bf \sigma_3}\quad,\quad
{\bf S_2}\equiv -\frac{g_0\,w_0}{2}\,{\bf\sigma_3}
\ee
By direct computation of (\ref{eq:QIR}), and using the fact that the eigenvalues of the matrices
${\bf S_\pm}\equiv {\bf S_1}\mp{\bf S_2}$ are real and given respectively by,
\be
-\frac{\omega}{s_0} \pm\sqrt{\left(\frac{\omega}{s_0}\right)^2 + \kappa_+{}^2}\qquad,\qquad
-\frac{\omega}{s_0} \pm\sqrt{\left(\frac{\omega}{s_0}\right)^2 + \kappa_-{}^2}
\ee
with $\kappa_\pm{}^2$ defined in (\ref{eq:kappa+-}), it is little tedious but straightforward to show that
$Q_{IR}(k)>0$ except when we are outside both IR Dirac cones, i.e. $\kappa_\pm{}^2 > 0$, in which case $Q_{IR}(k)=0$.
So, since current conservation implies $Q_{UV}(k) = Q_{IR}(k)=Q(k)$ we can conclude that effectively $\rho(k)\geq 0$,
being zero outside both IR Dirac cones or in other words, $\rho$ is strictly positive inside some IR Dirac cone
and null outside both of them.

When $T>0$ things are much simpler; the behavior for $z\rightarrow 1^-$ is of the form,
\be
\vec u_k(z)\simeq\exp^{-i\,\frac{\omega}{4\,\pi\,T}\,\ln(1-z)}\,\vec b
\label{eq:comportamiento-IR_ibc Sno0-app}
\ee
where $\vec b$ is read from (\ref{eq:comportamiento-IR_ibc Sno0}) in terms of $\vec u_k^{(ir)}$.
From (\ref{eq:Q}) we get,
\be
Q_{IR}(k) = 2\,\vec b{}^\dagger\,\vec b > 0
\ee
So we find that at finite temperature the spectral function must be strictly positive anywhere in space of momenta.
\section{On the spectral function in conformally flat backgrounds}
\label{sec:appendix-conformally-flat}
This appendix intends to clarify the relation among in-going b.c., $\omega$ prescriptions
and the positivity of the spectral function when arbitrary conformally flat backgrounds are considered.

We consider the following backgrounds, solution to the equations of motion (\ref{eq:eom-bosonic})
\bea
G&=&\frac{1}{z^2}\,\left( -dt^2 + d\vec x^2 + dz^2 \right)\cr
{\bf A} &=& \left( dt\,\phi + dx^1\,W_1 + dx^2\,W_2\right)\,\tau_0.
\eea
where $(\phi,\,W_1,\,W_2)$ are constant and we take the Cartan subalgebra in the $\tau_0$-direction.

Working in momentum-space,
$\Psi(x,z) = z^{3/2}\,e^{i\,k_\mu\, x^\mu}\left(\begin{array}{l}\psi^+_k(z)\\ \psi^-_k(z) \end{array}\right)$
and after introducing the notation (\ref{eq:four-tuples}), the equations of motion result as in (\ref{eq:eom-four-tuples})
with the constant matrix $\bf U$ defined by
\be
{\bf U} = \left(\begin{array}{cc}{\bf U_+}&{\bf 0}\\{\bf 0}&{\bf U_-}\end{array}\right)\qquad;\qquad
{\bf U_\pm}\equiv -\lambda_a^\pm\,\gamma^a
\ee
where
\bea
\lambda_0^\pm &=& -\omega \mp\frac{\phi}{2}\qquad,\qquad
\lambda_1^\pm = k_x \mp\frac{W_1}{2}\qquad,\qquad\lambda_2^\pm = k_y \mp\frac{W_2}{2}\cr
\lambda_\pm{}^2 &\equiv& \eta^{ab}\,\lambda_a^\pm\,\lambda_b^\pm =
-\left(\omega\pm\frac{\phi}{2}\right)^2 + \left(k_x \mp\frac{W_x}{2}\right)^2+ \left(k_y \mp\frac{W_y}{2}\right)^2
\eea
The general solution of (\ref{eq:eom-four-tuples}) is
\be
\vec u_k(z) = \left(\begin{array}{cc}
e^{\sqrt{\lambda_+{}^2}\,z}\;{\bf a^+} + e^{-\sqrt{\lambda_+{}^2}\,z}\;{\bf b^+}\\
e^{\sqrt{\lambda_-{}^2}\,z}\;{\bf a^-} + e^{-\sqrt{\lambda_-{}^2}\,z}\;{\bf b^-}\end{array}\right)\qquad;\qquad
\vec v_k(z) =  i\,\left(\begin{array}{cc}\frac{{\bf U_+}}{\lambda_+{}^2}&{\bf 0}\\
{\bf 0}&\frac{{\bf U_-}}{\lambda_-{}^2}\end{array}\right)\,\vec u'_k(z)
\ee
Let us define,
\be\label{eq:lambda2}
\sqrt{\lambda_\pm{}^2} = \left\{
\begin{array}{c}
\lambda_\pm \geq 0 \qquad\qquad\qquad,\quad \lambda_\pm{}^2\geq 0\\
i\,sign(\lambda_0^\pm)\,\sqrt{-\lambda_\pm{}^2} \quad,\quad \lambda_\pm{}^2< 0
\end{array}\right.
\ee
and take as b.c. ${\bf a^\pm}={\bf 0}$;
it is clear that in the first case it is just a smoothness condition when $z\rightarrow\infty$.
With this b.c. the solution can be written
\bea
\vec u_k(z) &=& \left(\begin{array}{cccc}
e^{-\sqrt{\lambda_+{}^2}\,z}\;{\bf 1}&{\bf 0}\\{\bf 0}&
e^{-\sqrt{\lambda_-{}^2}\,z}\;{\bf 1}\end{array}\right)\;\vec u_k^{(0)}\cr
\vec v_k(z) &=& -i\, \left(\begin{array}{cccc}
e^{-\sqrt{\lambda_+{}^2}\,z}\;{\bf 1}&{\bf 0}\\{\bf 0}&
e^{-\sqrt{\lambda_-{}^2}\,z}\;{\bf 1}\end{array}\right)\,
\left(\begin{array}{cc}\frac{{\bf U_+}}{\sqrt{\lambda_+{}^2}}&{\bf 0}\\
{\bf 0}&\frac{{\bf U_-}}{\sqrt{\lambda_-{}^2}}\end{array}\right)\;\vec u_k^{(0)}
\eea
From the definition (\ref{eq:M}) and the AdS/CFT recipe equation (\ref{eq:GMC}) we read the retarded Green function
\be
{\bf G_R}(k) = \left(\begin{array}{cccc}
-\frac{\lambda_a^+}{\sqrt{\lambda_+{}^2}}\;\gamma^a\,\gamma^0&{\bf 0}\\{\bf 0}&
-\frac{\lambda_a^-}{\sqrt{\lambda_-{}^2}}\;\gamma^a\,\gamma^0\end{array}\right)
\ee
The spectral function is then obtained from (\ref{eq:rho}),
\be\label{eq:CFBrho}
\rho(\omega, k_x,k_y) = 2\,\left(
\frac{|\omega+\frac{\phi}{2}|}{\sqrt{-\lambda_+{}^2}}\;\theta(-\lambda_+{}^2) + \frac{|\omega-\frac{\phi}{2}|}{\sqrt{-\lambda_-{}^2}}\;\theta(-\lambda_-{}^2)\right)
\ee
We see that the b.c. adopted for $z\rightarrow\infty$ together with the introduction in equation (\ref{eq:lambda2}) of the right signs
in the definition of the roots of $\lambda_\pm{}^2$ in the region where they are negative, i.e. inside the Dirac cones
(and where the spectral function is non-zero)  is what assure the positivity of the spectral function.
And when $\phi=0$, the relevant case in the IR limit at null entropy considered in the paper, they coincide with the ingoing b.c. in the
region $\kappa_\pm{}^2< 0$ considered in subsection \ref{sec:fermions-zero-entropy}.

As a last remark, it is easy to see that these b.c. effectively follow from the prescription of smoothness after shifting $\omega\rightarrow \omega + i\epsilon$ with $\epsilon\rightarrow 0^+$;  in fact if we assume that $\lambda_\pm{}^2< 0$
we have,
\be
\sqrt{\lambda_\pm{}^2}|_{\omega\rightarrow \omega + i\epsilon} \approx \sqrt{\lambda_\pm{}^2}
+ i \,\frac{\lambda_0^\pm}{\sqrt{\lambda_\pm{}^2}}\,\epsilon\quad\Longrightarrow\quad
e^{-\sqrt{\lambda_\pm{}^2}\,z}|_{\omega\rightarrow \omega + i\epsilon} \approx e^{-\sqrt{\lambda_\pm{}^2}\,z}\,
e^{-i\frac{\lambda_0^\pm}{\sqrt{\lambda_\pm{}^2}}\,\epsilon\,z}
\ee
and then the exponential will be well-behaved at $z\rightarrow\infty$  if (\ref{eq:lambda2}) holds \cite{Gubser:2010dm}.


\end{document}